\documentclass[12pt]{article}

\usepackage{graphicx}
\usepackage{float}
\usepackage{cite}
\usepackage{amsfonts}
\usepackage{amssymb}
\usepackage{amsmath}
\usepackage{relsize}
\usepackage[compatibility=false]{caption}
\usepackage{subcaption}
\usepackage{xcolor}

\def\gtwid{\mathrel{\raise.3ex\hbox{$>$\kern-.75em\lower1ex\hbox{$\sim$}}}}
\def\ltwid{\mathrel{\raise.3ex\hbox{$<$\kern-.75em\lower1ex\hbox{$\sim$}}}}
\def\square{\kern1pt\vbox{\hrule height 1.2pt\hbox{\vrule width 1.2pt\hskip 3pt
   \vbox{\vskip 6pt}\hskip 3pt\vrule width 0.6pt}\hrule height 0.6pt}\kern1pt}

\begin{document}

\begin{titlepage}

\begin{flushright}
UFIFT-QG-26-06
\end{flushright}

% \vskip 0.5cm

\begin{center}
{\bf Universal Secular External Leg Corrections for Gauge Independent Scalar 
Self-Mass on de Sitter}
\end{center}

% \vskip 0.25cm

\begin{center}
D. Glavan$^{1*}$, S. P. Miao$^{2\star}$, T. Prokopec$^{3\dagger}$ 
and R. P. Woodard$^{4\ddagger}$
\end{center}

\begin{center}
\it{$^{1}$ CEICO, FZU — Institute of Physics of the Czech Academy of Sciences, \\
Na Slovance 1999/2, 182 21 Prague 8, CZECH REPUBLIC}
\end{center}

\begin{center}
\it{$^{2}$ Department of Physics, National Cheng Kung University, \\
No. 1 University Road, Tainan City 70101, TAIWAN}
\end{center}

\begin{center}
\it{$^{3}$ ITP, EMMEPH and Spinoza Institute, Utrecht University, \\
Princetonplein 5, 3584 CC Utrecht, THE NETHERLANDS}
\end{center}

\begin{center}
\it{$^{4}$ Department of Physics, University of Florida,\\
Gainesville, FL 32611, UNITED STATES}
\end{center}

% \vspace{0.25cm}

\begin{center}
ABSTRACT
\end{center}
This work concerns a procedure for removing gauge dependence from
graviton corrections to the effective field equations for a massless,
minimally coupled scalar on de Sitter background. The procedure 
involves combining diagrams in the same way as the scattering amplitude 
for the $t$-channel exchange of the massless scalar between two massive 
particles, but {\it without} taking the asymptotic limits which are
problematic in cosmology. Implementing this at 1-loop on flat space 
background requires five classes of diagrams, in addition to the naive 
exchange, and the combination of those diagrams does eliminate dependence 
on the graviton gauge fixing functional. When those same diagrams are 
generalized to de Sitter background, their combination cancels the most 
important (``nonlocal'') secular gauge dependence, but it leaves 
(``local'') secular gauge dependence that is associated with external 
legs. Local secular gauge dependence can be removed by adding two 
additional classes of diagrams, and making secular corrections to the 
external wavefunctions. The purpose of this paper is to include these
additional corrections on de Sitter background in order to derive
fully gauge independent effective field equations at 1-loop order. 

\begin{flushleft}
\end{flushleft}

{\color{blue}
\begin{minipage}{2in}
$^{*}$ glavan@fzu.cz \\
$^{\dagger}$ T.Prokopec@uu.nl
\end{minipage}
\hfill
\begin{minipage}{2.3in}
$^{\star}$ spmiao5@mail.ncku.edu.tw \\
$^{\ddagger}$ woodard@phys.ufl.edu
\end{minipage}
}

\end{titlepage}

\section{Introduction}

From the perspective of cosmology one can see that the logical
structure of quantum field theory has been somewhat distorted
by its adaptation to the problem of scattering on flat space 
background. One of the ways this can manifest is the dismissal 
of the loop-corrected effective field equations. In classical 
electromagnetism one is happy to solve Maxwell's equations, 
and no one doubts the physical relevance of the resulting 
electric and magnetic fields at finite points in spacetime. 
Yet quantum field theory has accustomed us to accepting the 
unphysicality of local fields as soon as quantum corrections
are included, no matter how small they might be. We are instead 
instructed that the only valid conclusions derive from asymptotic 
scattering amplitudes. So rather than computing the 1-loop vacuum
polarization, then using it to quantum-correct Maxwell's equations,
and studying the response to a point charge to see the famous 
strengthening of the electrodynamic force at short distances, one
must instead compute the scattering amplitude for the exchange of
a photon and then use inverse scattering theory to infer the 
quantum correction to the Coulomb potential.

Traditional quantum field theory supplies three reasons for this
curious attitude:
\begin{enumerate}
\item{The effective field equations are not real, so that even
real charges and currents would engender complex electric and
magnetic fields;}
\item{The effective field equations are not causal, so that 
solutions depend on conditions outside the past light-cone; and}
\item{The effective field equations depend upon the gauge field
and graviton gauge in which the 1PI (one-particle-irreducible)
$n$-point functions are evaluated, so that solutions can be 
affected by arbitrary gauge choices.}
\end{enumerate}
The need to formulate quantum field theory on cosmological 
backgrounds has taught us that the first two objections derive
from too great a reliance on the in-out formalism which is 
appropriate to computing asymptotic scattering amplitudes. In the
in-out formalism, the 1PI $n$-point functions which comprise the 
effective field equation represent matrix elements between free 
vacuum at asymptically early and late times. That is why they are 
not real, even for Hermitian fields, and why they have support for
spacelike separations. 

The resolution to the first two problems is provided by the 
Schwinger-Keldysh formalism, a diagrammatic technique for computing 
true expectation values which is almost as simple to use as the 
in-out Feynman rules \cite{Schwinger:1960qe,Mahanthappa:1962ex,
Bakshi:1962dv,Bakshi:1963bn,Keldysh:1964ud} The effective field 
equations of the Schwinger-Keldysh formalism are real for Hermitian
fields. They are also causal, in the sense that the field at 
spacetime point $x^{\mu}$ is only influenced by points ${x'}^{\mu}$ 
on or within the past light-cone \cite{Chou:1984es,Jordan:1986ug,
Calzetta:1986ey}. Finally, it is trivial to convert the usual in-out 
effective field equations to those of the Schwinger-Keldysh 
formalsim \cite{Ford:2004wc}.

Resolving the third problem, gauge dependence, requires more work,
and is still an open problem. A procedure has been developed that 
works by including quantum gravitational correlations with the 
source which disturbs the effective field and the observer who 
detects it \cite{Miao:2017feh}. The procedure is to build the same 
diagrams that would go into an S-matrix element and then simplify 
them with a series of relations derived by Donoghue 
\cite{Donoghue:1993eb,Donoghue:1994dn,Donoghue:1996mt} to capture the 
$t$-channel exchange. In the end few traces of the source and observer 
remain, and each of the simplified diagrams can be regarded as a 
correction to the 1PI 2-point function in the linearized, effective 
field equation.

Consider graviton corrections to a massless, minimally coupled scalar 
on flat space background ($g_{\mu\nu} = \eta_{\mu\nu} + \kappa h_{\mu\nu}$
with $\kappa^2 = 16 \pi G$) in the 2-parameter family of covariant gauge 
fixing functions,
\begin{equation}
\mathcal{L}_{GF} = -\tfrac1{2 \alpha} \eta^{\mu\nu} F_{\mu} F_{\nu} \qquad , 
\qquad F_{\mu} = \eta^{\rho\sigma} \Bigl( h_{\mu \rho , \sigma} - 
\tfrac{\beta}{2} h_{\rho\sigma , \mu} \Bigr) \; . \label{gauge}
\end{equation}
\noindent In the Schwinger-Keldysh formalism, the renormalized self-mass is 
\cite{Miao:2017feh},
\begin{eqnarray}
M_0^2(x;x') & = & -\mathcal{C}_0(\alpha,\beta) \times 
\tfrac{\kappa^2 \partial^8}{2^7 \pi^3} \Bigl\{ \theta(\Delta t \!-\! \Delta r) 
\Bigl( \ln[\mu^2 (\Delta t^2 \!-\! \Delta r^2)] - 1\Bigr) \Bigr\} , 
\label{M0} \qquad \\
\mathcal{C}_0(\alpha,\beta) & = & +\tfrac34 -\tfrac34 \times \alpha - 
\tfrac32 \times \tfrac1{\beta - 2} + \tfrac34 \times 
\tfrac{(\alpha - 3)}{(\beta - 2)^2} \; , \label{C0}
\end{eqnarray}
where $\Delta t \equiv t - t'$ and $\Delta r \equiv \Vert \vec{x} - 
\vec{x}' \Vert$. The linearized effective field equation,
\begin{equation}
\partial^2 \Phi(x) - \int \!\! d^4x \, M_0^2(x;x') \Phi(x') = J(x) \; ,
\end{equation}
is real and causal, as advertised, but highly gauge-dependent.

\begin{figure}[ht]
\includegraphics[width=3.3cm]{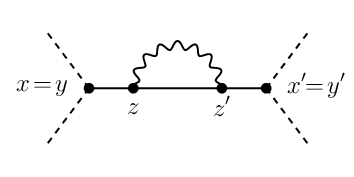} \hskip .2cm
\includegraphics[width=3.3cm]{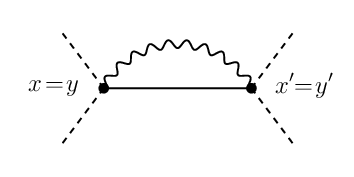}
\includegraphics[width=3.3cm]{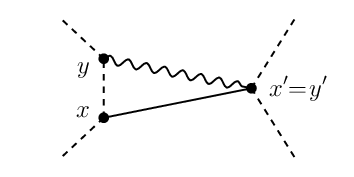}
\includegraphics[width=3.3cm]{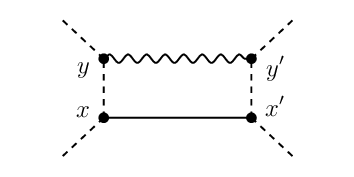}
\vskip -0.5cm
\caption{\footnotesize The left-hand diagram gives massive scalar (dashed 
lines) scattering from the self-mass (\ref{M0}), which is $i=0$ in the 
scheme of Table~\ref{Cab}. The second diagram includes graviton correlations 
between the vertices ($i=1$); the third diagram represents graviton 
correlations between the massive scalar propagator and a vertex ($i=2$), 
and the final diagram represents graviton correlations between the two 
massive scalars ($i=4$). Solid lines represent the massless scalar and wavy 
lines the graviton.}
\label{diagrams}
\end{figure}

Now, imagine quantum gravitational corrections to the scattering of 
two massive scalars by the exchange of such a massless scalar. 
Figure~\ref{diagrams} shows four of the many diagrams that contribute.
In position space these diagrams consist of products 
of (possibly differentiated) massive and massless propagators,
$i\Delta_m(x;x')$ and $i\Delta(x;x')$, respectively. All the 3-point 
and 4-point diagrams can be reduced to 2-point form by applying 
position-space versions of the Donoghue Identities \cite{Miao:2017feh,
Katuwal:2021thy}, of which the 3-point one is,
\begin{equation}
i\Delta_m(x;y) i\Delta(x;x') i\Delta(y;x') \longrightarrow
\tfrac{i\delta^D(x - y)}{2 m^2} \, [ i\Delta(x;x')]^2 \; . \label{IntID1}
\end{equation}
\noindent Any 2-point contribution so obtained can be regarded as a correction 
to the self-mass through an identity based on the massless propagator 
equation $\partial^2 i\Delta(x;x') = i\delta^D(x - x')$,
\begin{equation}
f(x;x') = -\int d^Dz \, i\Delta(x;z) \int d^Dz' \, i\Delta(x';z') \times
\partial_z^2 \partial_{z'}^2 f(z;z') \; . \label{keytrick}
\end{equation}
By Poincar\'e invariance, the function $\partial_x^2 \partial_{x'}^2 
f(x;x')$ takes the same form as (\ref{M0}), but with a different gauge-dependent
coefficient, $\mathcal{C}_i(\alpha,\beta)$. Table~\ref{Cab} collects the 
results, whose sum is indeed independent of $\alpha$ and $\beta$.
\begin{table}[H]
\setlength{\tabcolsep}{8pt}
\def\arraystretch{1.5}
\centering
\begin{tabular}{|@{\hskip 1mm }c@{\hskip 1mm }||c|c|c|c|c|}
\hline
$i$ & $1$ & $\alpha$ & $\frac1{\beta-2}$ & $\frac{(\alpha-3)}{(\beta-2)^2}$ &
{\rm Description} \\
\hline\hline
0 & $+\frac34$ & $-\frac34$ & $-\frac32$ & $+\frac34$ &
{\rm scalar\ exchange} \\
\hline
1 & $0$ & $0$ & $0$ & $+1$ & {\rm vertex-vertex} \\
\hline
2 & $0$ & $0$ & $0$ & $0$ & {\rm vertex-source,observer} \\
\hline
3 & $0$ & $0$ & $+3$ & $-2$ & {\rm vertex-scalar} \\
\hline
4 & $+\frac{17}4$ & $-\frac34$ & $0$ & $-\frac14$ &
{\rm source-observer} \\
\hline
5 & $-2$ & $+\frac32$ & $-\frac32$ & $+\frac12$ &
{\rm scalar-source,observer} \\
\hline\hline
Total & $+3$ & $0$ & $0$ & $0$ & \\
\hline
\end{tabular}
\caption{\footnotesize The gauge-dependent factors $C_i(\alpha,\beta)$ 
for each contribution to the invariant scalar self-mass-squared. Table
taken from \cite{Miao:2017feh}.}
\label{Cab}
\end{table}

The same procedure can be used to remove gauge dependence from the
linearized effective field equation of any massless field. The steps are: 
\begin{enumerate}
\item{Write down (in position space) the diagrams which contribute to
quantum gravitational corrections to the scattering of two massive 
particles through the exchange of the massless field in question;}
\item{Reduce the 3-point and 4-point diagrams to 2-point form using
the Donoghue Identities; and}
\item{Regard each 2-point contribution as a correction to the 1PI 2-point
function of the massless field by the relation (\ref{keytrick}).}
\end{enumerate}
This has been done on flat space background for quantum gravitational
corrections to Maxwell's equation, and explicit calculation again shows
that the result is independent of the graviton gauge-fixing parameters 
$\alpha$ and $\beta$ \cite{Katuwal:2021thy}.

The de Sitter generalization of this computation involves reducing the
contributing diagrams to the scalar self-mass and showing that the sum
does not depend on the gauge-fixing parameters $\alpha$ and $\beta$ in
the de Sitter generalization of (\ref{gauge}),
\begin{equation}
\mathcal{L}_{\rm GF} = -\tfrac1{2\alpha} a^{D-2} \eta^{\mu\nu} F_{\mu} 
F^{\nu} \quad , \quad F_{\mu} = \eta^{\rho\sigma} \Bigl( h_{\mu \rho , 
\sigma}\ \!-\! \tfrac{\beta}2 h_{\rho\sigma , \mu} \!+\! (D\!-\!2) a H 
h_{\mu\rho} \delta^0_{~\sigma} \Bigr) \; . \label{DSgauge}
\end{equation}
The graviton propagator in this family of gauges is available for 
$\alpha = 1 + \delta \alpha$ and $\beta = 1 + \delta \beta$, to first
order in $\delta \alpha$ and $\delta \beta$ \cite{Glavan:2019msf},
which should be sufficient to establish gauge dependence. The computation
for $\alpha = \beta = 1$ has been made \cite{Glavan:2021adm,Glavan:2024elz}.
The dependence on the gauge-fixing parameter $\alpha$, for which we know
the full propagator \cite{Glavan:2025azq}, has been checked 
\cite{Glavan:2026pug}. The result is that the de Sitter 
generalizations of the Diagrams 1-5 cancel the most important 
(``nonlocal'') secular gauge dependence, but they leave a (``local'') 
secular gauge dependence which is associated with external legs. Local 
secular gauge dependence can be removed by adding two additional classes 
of diagrams, and making secular corrections to the external wavefunctions. 
The purpose of this paper is to include these additional corrections on 
de Sitter background in order to derive fully gauge independent effective 
field equations at 1-loop order. Among other things, this establishes 
the reality of graviton-induced secular logarithms in the scalar 
self-mass, which is all we consider here. These secular logarithms 
ultimately induce a large spatial logarithm \cite{Glavan:2021adm,
Glavan:2024elz} in the 1-loop correction to the classical potential of a 
point scalar charge \cite{Burko:2002ge,Akhmedov:2010ah,Glavan:2019yfc}.

This paper has 6 sections, of which the first is this Introduction. In
section 2 we review the original de Sitter calculation of diagrams 0
through 5 \cite{Glavan:2021adm,Glavan:2024elz}. Then section 3 and 4
give two additional diagrams that do not contribute on flat space 
background but do contribute on de Sitter \cite{Glavan:2026pug}. Section 
5 derives the secular part of the external wave function correction,
which is absent in flat space but not on de Sitter \cite{Glavan:2026pug}. 
Our conclusions comprise section 6.

\section{Diagrams 0 through 5}

The purpose of this section is to review the de Sitter generalization 
of the six flat space diagrams presented in Table~\ref{Cab}. We begin 
by presenting the Feynman rules. Then a discussion is given of the 
close connection between large logarithms and renormalization for the 
sort of graviton loop corrections to scalars which are the subject 
of this work. The section concludes by presenting the six diagrams and
given the final results for useful combinations of them.

\subsection{Feynman Rules}

The basic system we seek to study is a massless scalar $\phi$ that is
minimally coupled to the metric $g_{\mu\nu} = a^2 (\eta_{\mu\nu} +
\kappa h_{\mu\nu})$, where $\kappa^2 \equiv 16 \pi G$ is the quantum
gravitational loop counting parameter. We work on $D$-dimensional de 
Sitter background, which corresponds to cosmological scale factor
$a(\eta) \equiv -\frac1{H \eta}$, where $\eta$ is the conformal time
and $H^2 \equiv (D-1) \Lambda$. To this system we add a massive, 
minimally coupled scalar $\psi$, which serves as the source who excites 
the massless scalar, and as the observer who measures it. The complete 
Lagrangian is,
\begin{eqnarray}
\lefteqn{\mathcal{L} = \tfrac{R - (D-2) \Lambda}{16 \pi G} - \tfrac12 
\partial_{\mu} \phi \partial_{\nu} \phi g^{\mu\nu} \sqrt{-g} } 
\nonumber \\
& & \hspace{3cm} - \tfrac12 \partial_{\mu} \psi \partial_{\nu} \psi 
g^{\mu\nu} \sqrt{-g} -\tfrac12 m^2 \psi^2 \sqrt{-g} - \tfrac12 \lambda 
\phi \psi^2 \sqrt{-g} \; . \label{Lagrangian} \qquad
\end{eqnarray} 
The basic system involving $\phi$ and $h_{\mu\nu}$ is on the first line;
parts involving the source and observer field $\psi$ are on the second
line.

We use Bunch-Davies vacuum, which corresponds to a state which is
empty of particles in the distant past \cite{Chernikov:1968zm,
Schomblond:1976xc,Bunch:1978yq}. Propagators depend principally on 
the de Sitter length function,
\begin{equation}
y(x;x') \equiv a a' H^2 \Delta x^2 \qquad , \qquad \Delta x^2 \equiv
\Vert \vec{x} \!-\! \vec{x}'\Vert^2 - (\vert \eta \!-\! \eta'\vert 
\!-\! i \epsilon)^2 \; . \label{ydef}
\end{equation}
The massive scalar propagator is \cite{Chernikov:1968zm},
\begin{eqnarray}
\lefteqn{i\Delta_m(x;x') = \tfrac{H^{D-2}}{(4\pi)^{\frac{D}2}} 
\Bigl\{ \tfrac{\Gamma(\frac{D}2)}{\frac{D}2 - 1} (\tfrac{4}{y})^{
\frac{D}2 - 1} + \tfrac{\Gamma(\frac{D}{2} - 1) \Gamma(2 - \frac{D}2)}{
\Gamma(\frac12 + i \mu) \Gamma(\frac12 - i \mu)} } \nonumber \\
& & \hspace{0.5cm} \times \!\! \sum_{n=0}^{\infty} \Bigl[ \tfrac{
\Gamma(\frac32 + n + i \mu) \Gamma(\frac32 + n - i\mu)}{\Gamma(3 - 
\frac{D}2 + n) (n+1)!} (\tfrac{y}{4})^{n - \frac{D}2 + 2} \!\!-\!
\tfrac{\Gamma(\frac{D-1}{2} + n + i\mu) \Gamma(\frac{D-1}{2} + n 
- i\mu)}{\Gamma(\frac{D}{2} + n) n!} (\tfrac{y}{4})^n \Bigr] \Bigr\} , 
\qquad \label{DeltaM}
\end{eqnarray}
where the index is $\mu \equiv \sqrt{\frac{m^2}{H^2} - 
(\frac{D-1}{2})^2}$. The propagator obeys,
\begin{equation}
(\mathcal{D} - m^2 a^D) i\Delta_m(x;x') = i \delta^D(x \!-\! x') 
\qquad , \qquad \mathcal{D} \equiv \partial_{\mu} a^{D-2} \partial^{\mu}
\; . \label{Ddef}
\end{equation}
Each of the four external legs is nominally integrated up against a
massive scalar mode function or its complex conjugate,
\begin{equation}
u(x;\vec{k}) = \tfrac{\exp[-i \mu H t + i \vec{k} \cdot \vec{x}]}{\sqrt{2 
\mu H a^{D-1}}} \sum_{n=0}^{\infty} \tfrac{\Gamma(1 + i \mu)}{n! \Gamma(1
+ n + i \mu)} (-\tfrac{k^2}{4 H^2 a^2})^n \qquad , \qquad t \equiv
\tfrac1{H} \ln(a) \; . \label{udef}
\end{equation}

In contrast to its massive cousin, the masseless, minimally coupled 
scalar propagator breaks de Sitter invariance \cite{Allen:1987tz,
Onemli:2002hr},
\begin{eqnarray}
\lefteqn{i\Delta_{A}(x;x') = \tfrac{H^{D-2}}{(4\pi)^{\frac{D}2}}
\Bigl\{ \tfrac{\Gamma(\frac{D}2)}{\frac{D}2 - 1} (\tfrac{4}{y})^{
\frac{D}2 -1} + \tfrac{\Gamma(\frac{D}2 +1)}{\frac{D}2 -2} 
(\tfrac{4}{y})^{\frac{D}2 -2} \Bigr\} + k \Bigl\{ \ln(a a') + 
\Psi_{A}\Bigr\} } \nonumber \\
& & \hspace{2.5cm} - \tfrac{H^{D-2}}{(4\pi)^{\frac{D}2}} 
\sum_{n=1}^{\infty} \Bigl[ \tfrac{\Gamma(\frac{D}2 +1 + n)}{(n 
-\frac{D}2 +2) (n+1)!} (\tfrac{y}{4})^{n-\frac{D}2 +2} -
\tfrac{\Gamma(D-1+n)}{n \Gamma(\frac{D}2 +n)} (\tfrac{y}{4})^n \Bigr]
\; . \qquad \label{DeltaA} 
\end{eqnarray}
The constants $k$ and $\Psi_A$ are,
\begin{eqnarray}
k &\!\!\! \equiv \!\!\!& \tfrac{H^{D-2}}{(4\pi)^{\frac{D}2}} 
\tfrac{\Gamma(D-1)}{\Gamma(\frac{D}2)} \; , \qquad \label{kdef} \\
\Psi_{A} &\!\!\! \equiv \!\!\!& -\psi(1 \!-\! \tfrac{D}2) + 
\psi(\tfrac{D-1}{2}) + \psi(D \!-\! 1) - \gamma \; . \qquad \label{Psidef}
\end{eqnarray}
The propagator equation for $i\Delta_{A}$ is,
\begin{equation}
\mathcal{D} i\Delta_A(x;x') = i\delta^D(x \!-\! x') \; . \label{DAeqn}
\end{equation}
Near coincidence (${x'}^{\mu} \rightarrow x^{\mu}$), the leading 
singularities of (\ref{DeltaM}) and (\ref{DeltaA}) are the same,
\begin{eqnarray}
i\Delta_m(x;x') &\!\!\! = \!\!\!& \tfrac{\Gamma(\frac{D}2 -1)}{4 
\pi^{\frac{D}2}} \tfrac1{[a a' \Delta x^2]^{\frac{D}2 -1}} + 
O(\Delta x^{4-D}) \; , \label{leadingM} \\
i\Delta_A(x;x') &\!\!\! = \!\!\!& \tfrac{\Gamma(\frac{D}2 -1)}{4 
\pi^{\frac{D}2}} \tfrac1{[a a' \Delta x^2]^{\frac{D}2 -1}} + 
O(\Delta x^{4-D}) \; . \label{leadingA}
\end{eqnarray}
The big difference is that, for $D=4$, $i\Delta_m$ still has an infinite
series expansion, the series expansion of $i\Delta_A$ terminates,
\begin{equation}
D=4 \qquad \Longrightarrow \qquad i\Delta_A(x;x') = \tfrac{1}{4 \pi^2 a a' 
\Delta x^2} - \tfrac{H^2}{8 \pi^2} \Bigl[ \ln(H^2 \Delta x^2) \!+\! 2 \gamma
\!-\! 2\Bigr] \; . \label{4DDA}
\end{equation} 

The graviton propagator requires gauge fixing. The simplest gauge fixing
function \cite{Tsamis:1992xa,Woodard:2004ut} corresponds to setting $\alpha 
= \beta = 1$ in expression (\ref{DSgauge}). The two simple things about 
this gauge are:
\begin{enumerate}
\item{The graviton propagator consists of three scalar propagators, one 
of which is $i\Delta_A$, multiplying spacetime constant tensors; and}
\item{In $D=4$ dimensions, $i\Delta_A$ has only two terms (\ref{4DDA}), 
whereas the other two scalar propagators have only one term.}
\end{enumerate}
Then other two scalar propagators are,
\begin{eqnarray}
\lefteqn{i\Delta_{B} = \tfrac{H^{D-2}}{(4\pi)^{\frac{D}2}} \Bigl\{ 
\tfrac{\Gamma(\frac{D}2)}{\frac{D}2 -1} (\tfrac{4}{y})^{\frac{D}2 -1}
+ \sum_{n=0}^{\infty} \Bigl[ \tfrac{\Gamma(\frac{D}2 + n)}{(n+1)!} 
(\tfrac{y}{4})^{n-\frac{D}2 + 2} - \tfrac{\Gamma(D-2+n)}{\Gamma(\frac{D}2 +n)} 
(\tfrac{y}{4})^n \Bigr] \Bigr\} , } \label{DeltaB} \\
\lefteqn{i\Delta_{C} = \tfrac{H^{D-2}}{(4\pi)^{\frac{D}2}} \Bigl\{ 
\tfrac{\Gamma(\frac{D}2)}{\frac{D}2 -1} (\tfrac{4}{y})^{\frac{D}2 -1}
- \sum_{n=0}^{\infty} \Bigl[ \tfrac{(n-\frac{D}2+3) \Gamma(\frac{D}2 -1 + n)}{
(n+1)!} (\tfrac{y}{4})^{n-\frac{D}2 + 2} } \nonumber \\
& & \hspace{8cm} - \tfrac{(n + 1) \Gamma(D-3+n)}{\Gamma(\frac{D}2+n)}
(\tfrac{y}{4})^n \Bigr] \Bigr\} \; . \qquad \label{DeltaC}
\end{eqnarray}
They correspond to masses of $M^2_B = (D-2) H^2$ and $M_C^2 = 2 (D-3) H^2$,
respectively, which means they obey the equations,
\begin{equation}
[\mathcal{D} \!-\! (D\!-\!2) H^2 a^D] i\Delta_{B}(x;x') = i\delta^D(x \!-\! x')
= [\mathcal{D} \!-\! 2 (D\!-\!3) H^2 a^D] i\Delta_C(x;x') \; . \label{DBDCeqns}
\end{equation}
As stated, their $D=4$ expansions contain just one term,
\begin{equation}
D=4 \qquad \Longrightarrow \qquad i\Delta_B = i\Delta_C = \tfrac1{4 \pi^2 
a a' \Delta x^2} \; . \label{4DDBDC}
\end{equation}
In the gauge (\ref{DSgauge}) the graviton propagator is \cite{Tsamis:1992xa,
Woodard:2004ut},
\begin{eqnarray}
\lefteqn{i [\mbox{}_{\mu\nu} \Delta_{\rho\sigma}](x;x') = \Bigl[2 
\eta_{\mu (\rho} \eta_{\sigma )\nu} - \tfrac{2}{D-2} \eta_{\mu\nu} 
\eta_{\rho\sigma} \Bigr] \!\times\! i\Delta_C(x;x') } \nonumber \\
& & \hspace{-0.2cm} + \Bigl[2 \overline{\eta}_{\mu (\rho} 
\overline{\eta}_{\sigma )\nu} \!-\! \tfrac{2}{D-3} \overline{\eta}_{\mu\nu} 
\overline{\eta}_{\rho\sigma}\Bigr] \!\times\! i\Delta_{AC}(x;x') \!-\! 4 
\delta^{0}_{~(\mu} \overline{\eta}_{\nu)(\rho} \delta^0_{~\sigma)} 
\!\times\! i\Delta_{BC}(x;x') \; , \label{gravprop} \qquad 
\end{eqnarray}
where $i\Delta_{AC} \equiv i\Delta_{A} - i\Delta_{C}$, $i\Delta_{BC} \equiv
i\Delta_{B} - i\Delta_{C}$, parenthesized indices are symmetrized and 
$\overline{\eta}_{\mu\nu} \equiv \eta_{\mu\nu} + \delta^0_{~\mu} 
\delta^0_{~\nu}$ is the spatial part of the Minkowski metric. Note that the 
propagator differences $i\Delta_{AC}$ and $i\Delta_{BC}$ on the second 
line of (\ref{gravprop}) are less singular near coincidence because the most 
singular part of all three propagators (\ref{DeltaA}) and 
(\ref{DeltaB}-\ref{DeltaC}) is the same. 

The primitive contributions we derive in this paper require four vertices,
\begin{eqnarray}
\phi \psi^2 &\!\!\! \Longrightarrow \!\!\!& -\tfrac{i\lambda}{2} a^D \phi 
\psi^2 \; , \label{phi2psi} \qquad \\
\phi h \psi^2 &\!\!\! \Longrightarrow \!\!\!& -\tfrac{i\kappa \lambda}{4}
a^D \phi h \psi^2 \; , \label{hphi2psi} \qquad \\
h \psi^2 &\!\!\! \Longrightarrow \!\!\!& \tfrac{i\kappa}{2} a^{D-2} 
h^{\mu\nu} \partial_{\mu} \psi \partial_{\nu} \psi - \tfrac{i\kappa}{4} h 
[a^{D-2} \partial_{\rho} \psi \partial^{\rho} \psi \!+\! m^2 a^D \psi^2] 
\; , \label{h2psi} \\
h^2 \psi^2 &\!\!\! \Longrightarrow \!\!\!& \tfrac{i\kappa^2}{2} a^{D-2} 
[\tfrac12 h h^{\mu\nu} \!-\! h^{\mu \rho} h^{\nu}_{~\rho}] \partial_{\mu} 
\psi \partial_{\nu} \psi \nonumber \\
& & \hspace{3cm} - \tfrac{i \kappa^2}{2} [\tfrac18 h^2 \!-\! \tfrac14 
h^{\mu\nu} h_{\mu\nu}] [a^{D-2} \partial_{\rho} \psi \partial^{\rho} \psi
\!+\! a^D m^2 \psi^2] . \qquad \label{2h2psi}
\end{eqnarray}
The 3-point vertex (\ref{h2psi}) involves two derivatives of the scalar, 
one of which will act on an external mode function, with the other acting
on an internal scalar propagator. We employ a shorthand notation in which
an overline indicates the external derivative and a normal derivative acts
on the internal propagator. Momentum conservation at this vertex reads,
\begin{equation}
\overline{\partial}^{\mu} + \partial^{\mu} = -\widetilde{\partial}^{\mu}
- \partial_{a}^{\mu} \; , \label{conservation}
\end{equation}
where a tilde indicates that a derivative acting on the graviton propagator
and the subscript $a$ indicates a derivative acting on the factor of 
$a^{D-2}$. Many relations follow from (\ref{conservation}), including one
of great relevance to the final term of (\ref{h2psi}) \cite{Glavan:2024elz},
\begin{equation}
a^{D-2} \overline{\partial} \!\cdot\! \partial \!+\! m^2 a^D = \tfrac12 
\widetilde{\mathcal{D}} - \tfrac12 (\mathcal{D} \!-\! m^2 a^D) \; . 
\label{dbar.d}
\end{equation}
(Note that a term has been dropped because it vanishes when acted on the 
external state.) Three related identities are useful in reducing 4-point 
contributions,
\begin{eqnarray}
\lefteqn{\partial_{\mu} \Bigl[ f(x) a^{D-2} \partial^{\mu} \delta^D(x \!-\! x')
\Bigr] = \tfrac12 \mathcal{D} \Bigl[ f(x) \delta^D(x \!-\! x') \Bigr] } 
\nonumber \\
& & \hspace{4cm} + \tfrac12 \mathcal{D}' \Bigl[f(x) \delta^D(x \!-\! x')
\Bigr] \!-\! \tfrac12 \mathcal{D} f(x) \!\times\! \delta^D(x \!-\! x') \; , 
\qquad \label{4ptID1} \\
\lefteqn{a^{D-2} \nabla^2 \delta^D(x \!-\! x') = \mathcal{D} \Bigl[
\tfrac1{2 a H} \partial_0 + (\tfrac{D-1}{4})\Bigr] \delta^D(x \!-\! x') }
\nonumber \\
& & \hspace{6cm} + \mathcal{D}' \Bigl[ \tfrac1{2 a' H} \partial_0' + 
(\tfrac{D-1}{4})\Bigr] \delta^D(x \!-\! x') \; , \qquad \label{4ptID2} \\
\lefteqn{\ln(a) a^{D-2} \nabla^2 \delta^D(x \!-\! x') = \mathcal{D} \Bigl[
\tfrac{[\ln(a) + 1]}{2 a H} \partial_0 + (\tfrac{D-1}{4}) \ln(a) +
(\tfrac{D-2}{4})\Bigr] \delta^D(x \!-\! x') } \nonumber \\
& & \hspace{-0.7cm} + \mathcal{D}' \Bigl[\! \tfrac{[\ln(a') + 1]}{2 a' H} 
\partial_0' \!\!+\! (\tfrac{D-1}{4}) \ln(a') \!+\! (\tfrac{D-2}{4}) \!\Bigr] 
\delta^D(x \!-\! x') \!+\! (\tfrac{D-1}{2})^2 H^2 a^D \delta^D(x \!-\! x')
\; . \qquad \label{4ptID3}
\end{eqnarray}

\subsection{The Origin of Secular Logarithms}

Graviton-induced logarithms have two potential origins \cite{Woodard:2025smz}:
\begin{itemize}
\item{From the ``tail'' term of the propagator $i\Delta_A$ which is
evident in the order $H^2$ part (\ref{4DDA}) of its $D=4$ limit; and}
\item{From the mismatch between primitive divergences and the counterterms
that cancel them.}
\end{itemize}
It turns out that graviton loop corrections to the scalar self-mass being
studied are all of the second type \cite{Glavan:2021adm}. 

The mismatch arises because the $D$-dependent scale factors from vertices 
(\ref{phi2psi}-\ref{2h2psi}) are canceled by propagators. As an example,
consider the contribution to the massless, minimally coupled scalar self-mass
that comes from a loop of massive scalars interacting through two of the
$\phi \psi^2$ vertices (\ref{phi2psi}),
\begin{eqnarray} 
\lefteqn{\tfrac12 (-i \lambda a^D) \!\times\! [i \Delta_m(x;x')]^2 \!\times\! 
(-i \lambda {a'}^D) } \nonumber \\
& & \hspace{1.8cm} = -\tfrac{\lambda^2}{2} (a a')^D \!\times\! \Bigl\{ 
\tfrac{\Gamma^2(\frac{D}2 - 1)}{16 \pi^D} \tfrac1{[a a' \Delta x^2]^{D-2}} + 
{\rm Finite} \Bigr\} \; , \qquad \\
& & \hspace{1.8cm} = -\tfrac{\lambda^2 \Gamma(\frac{D}2 - 1)}{8 \pi^{\frac{D}2}}
\tfrac{\mu^{D-4} a^4 i\delta^D(x - x')}{2 (D-3) (D-4)} + \tfrac{\lambda^2 (a a')^2 
\partial^2}{128 \pi^4} [\tfrac{\ln(\mu^2 \Delta x^2)}{\Delta x^2}] + {\rm Finite} 
\; . \qquad \label{EG1}
\end{eqnarray}
The primitive divergence in expression (\ref{EG1}) could be canceled by
a 2-point counterterm,
\begin{equation}
\Delta \mathcal{L} = -\tfrac12 c \phi^2 \sqrt{-g} \qquad , \qquad c = -
\tfrac{\lambda^2 \Gamma(\frac{D}2 - 1)}{8 \pi^{\frac{D}2}} \tfrac{\mu^{D-4}}{
2 (D-3) (D-4)} \; . \label{DLscalar}
\end{equation}
Hence the primitive divergence goes like $a^4$, while the counterterm that
cancels it goes like $a^D$. The result is a finite temporal logarithm,
\begin{equation}
-\tfrac{\lambda^2 \Gamma(\frac{D}2 - 1)}{8 \pi^{\frac{D}2}}
\tfrac{\mu^{D-4} a^4 i\delta^D(x - x')}{2 (D-3) (D-4)} + 
\tfrac{\lambda^2 \Gamma(\frac{D}2 - 1)}{8 \pi^{\frac{D}2}}
\tfrac{\mu^{D-4} a^D i\delta^D(x - x')}{2 (D-3) (D-4)} = 
\tfrac{\lambda^2  \ln(a) a^4}{16 \pi^2} + O({\scriptstyle D - 4}) \; . 
\label{EG2}
\end{equation}

The fact that all the secular logarithms of interest for our problem have
their origin in this mismatch has a crucially important consequence. It 
means that {\it we can read the secular logarithms from the relatively 
simple primitive divergences}. We do not need to bother with the much more
complicated finite parts.

\subsection{Primitive Results for Diagrams $0$-$5$}

\begin{figure}[H]
\centering
\includegraphics[width=4cm]{diagram01.eps}
\includegraphics[width=4cm]{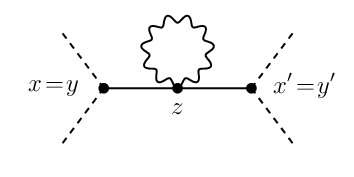}
\caption{\footnotesize Diagrams $0a$ (left) and $0b$ (right) included in the 
uncorrected, self-mass.}
\label{Diagram0}
\end{figure}

\begin{figure}[H]
\centering
\includegraphics[width=4cm]{diagram11.eps}
\includegraphics[width=4cm]{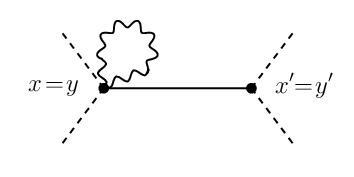}
\includegraphics[width=4cm]{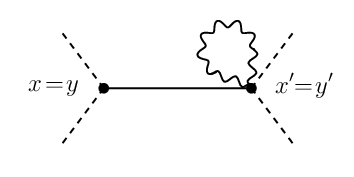}
\caption{\footnotesize Diagrams $1a$ (left), $1b$ (middle) and $1c$ (right) in 
which the vertex is corrected.}
\label{Diagram1}
\end{figure}

\begin{figure}[H]
\centering
\includegraphics[width=3.3cm]{diagram21.eps}
\includegraphics[width=3.3cm]{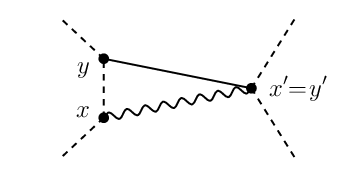}
\includegraphics[width=3.3cm]{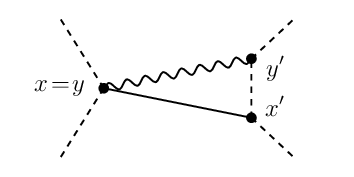}
\includegraphics[width=3.3cm]{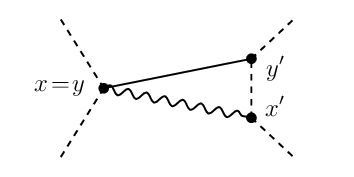}
\caption{\footnotesize Diagrams $2a$ (left), $2b$ (2nd), $2c$ (3rd) 
and $2d$ (right) including a correlation from the propagation of the source 
or observer with the vertex.}
\label{Diagram2}
\end{figure}

\begin{figure}[H]
\centering
\includegraphics[width=4cm]{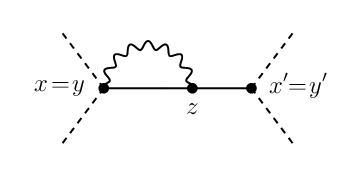}
\includegraphics[width=4cm]{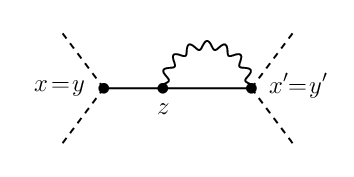}
\caption{\footnotesize Diagrams $3a$ (left) and $3b$ (right) which include 
a correlation between a vertex and the propagation of the exchange scalar.}
\label{Diagram3}
\end{figure}

\begin{figure}[H]
\centering
\includegraphics[width=3.3cm]{diagram41.eps}
\includegraphics[width=3.3cm]{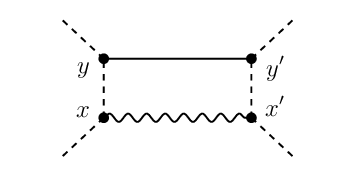}
\includegraphics[width=3.3cm]{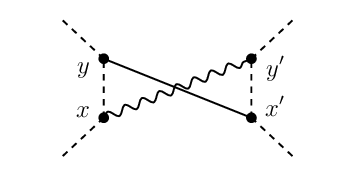}
\includegraphics[width=3.3cm]{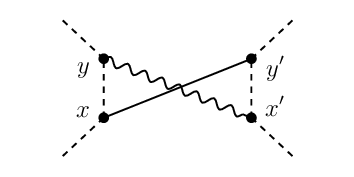}
\caption{\footnotesize Diagrams $4a$ (left), $4b$ (2nd), $4c$ (3rd)
and $4d$ (right) including a correlation between the propagation
of the source and the propagation of the observer.}
\label{Diagram4}
\end{figure}

\begin{figure}[H]
\centering
\includegraphics[width=3.3cm]{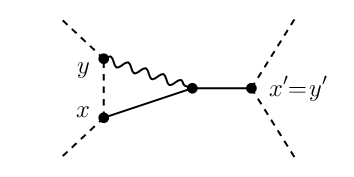}
\includegraphics[width=3.3cm]{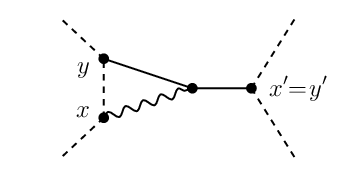}
\includegraphics[width=3.3cm]{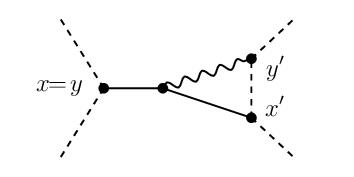}
\includegraphics[width=3.3cm]{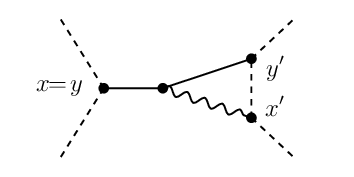}
\caption{\footnotesize Diagrams $5a$ (left), $5b$ (2nd), $5c$ (3rd) and
$5d$ (right) including a correlation between the propagation of the 
source or observer and the propagation of the exchange scalar.}
\label{Diagram5}
\end{figure}

The basic process we study is the $t$-channel scattering of two massive 
scalars through the mediation of a massless scalar. The six, single-graviton 
loop corrections to this that matter on flat space background are mentioned
in Table~\ref{Cab}, and their topologies are the same for de Sitter as for
flat space. The $i=0$ (scalar exchange) diagrams of Figure~\ref{Diagram0} 
correspond to the ``vacuum polarization'' of a graviton correction to the 
massless scalar propagator. Its actual form is complicated 
\cite{Glavan:2021adm} but the large logarithms it contributes are,
\begin{equation}
-i M^2_{0}(x;x') = -\tfrac{\kappa^2 H^2}{4 \pi^2} \partial_{x}^{\mu}
[ a_x^2 \ln(a_x) \partial^x_{\mu} i\delta^4(x \!-\! x')] + 
\tfrac{\kappa^2 H^2}{2 \pi^2} \partial_0^x [a_x^2 \ln(a_x) \partial^x_0
i\delta^4(x \!-\! x')] \; . \label{M0logs}
\end{equation}
\noindent The $i=1$ (vertex-vertex) diagrams of Figure~\ref{Diagram1} arise 
from correlations between gravitons in the interaction vertices between the 
massless and massive scalars.
\noindent The $i=2$ (vertex-source and vertex-observer) diagrams of
Figure~\ref{Diagram2} derive from graviton correlations from the propagation
of one of the massive scalar to the vertex of the other.
\noindent The $i=3$ (vertex-scalar) diagrams of Figure~\ref{Diagram3} 
represent correlations between gravitons in the massive scalar vertices
and the propagation of the massless exchange scalar.
\noindent The $i=4$ (source-observer) diagrams of Figure~\ref{Diagram4}
come from correlations between gravitons from the propagation of one
massive scalar to the other.
\noindent The $i=5$ (scalar-source and scalar-observer) diagrams of
Figure~\ref{Diagram5} correspond to graviton correlations between the
propagation of a massive scalar and the propagation of the massless
exchange scalar.

It turns out that Diagrams 1a, 2 and 4 can be usefully combined into a
form we call $\overline{\overline{4}}$. The large logarithms contributed
by this class of diagrams are \cite{Glavan:2024elz},
\begin{eqnarray}
\lefteqn{-i M^2_{\overline{\overline{4}}}(x;x') = -\tfrac{31 \kappa^2 
H^2}{8 \pi^2} \Bigl\{\mathcal{D}_{x} [\ln(a_x) i\delta^4(x \!-\! x')] 
\!+\! \mathcal{D}_{x'} [\ln(a_{x'}) i\delta^4(x \!-\! x')] \Bigr\} } 
\nonumber \\
& & \hspace{6.5cm} + \tfrac{\kappa^2}{4 \pi^2} \mathcal{D}_{x} 
\mathcal{D}_{x'} \Bigl\{ \tfrac{\ln(a_x) i\delta^4(x - x')}{a_x^4} 
\Bigr\} \; . \qquad \label{M4bar}
\end{eqnarray}
One can similarly combine Diagrams 3 and 5 to give a form we call 
$\overline{5}$. The large logarithms it contributes are \cite{Glavan:2024elz},
\begin{eqnarray}
\lefteqn{-i M^2_{\overline{5}}(x;x') = \tfrac{\kappa^2 H^2}{8 \pi^2} 
\Bigl\{\mathcal{D}_{x} [\ln(a_x) i\delta^4(x \!-\! x')] \!+\! 
\mathcal{D}_{x'} [\ln(a_{x'}) i\delta^4(x \!-\! x')] \Bigr\} } 
\nonumber \\
& & \hspace{6.5cm} + \tfrac{\kappa^2}{4 \pi^2} \mathcal{D}_{x} 
\mathcal{D}_{x'} \Bigl\{ \tfrac{\ln(a_x) i\delta^4(x - x')}{a_x^4} 
\Bigr\} \; . \qquad \label{M5bar}
\end{eqnarray}
The remaining parts of Diagram 1 contribute \cite{Glavan:2024elz},
\begin{equation}
-i M^2_{1bc}(x;x') = \tfrac{3 \kappa^2 H^2}{4 \pi^2} \Bigl\{\mathcal{D}_{x}
[\ln(a_x) i\delta^4(x \!-\! x')] \!+\! \mathcal{D}_{x'} [\ln(a_{x'}) 
i\delta^4(x \!-\! x')] \Bigr\} \; . \label{M1bc}
\end{equation}

\section{Diagram 6}

\begin{figure}[ht]
\centering
\includegraphics[width=12cm]{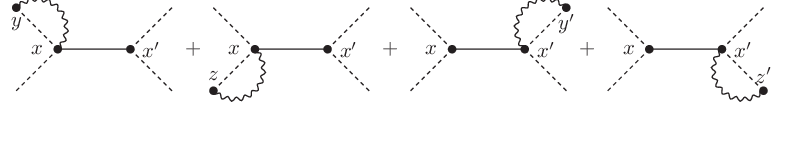}
\vskip -1cm
\caption{\footnotesize Diagram 6 correction. Solid lines
correspond to the massless scalar, dashed lines correspond to the
massive scalar, and wavy lines are gravitons.}
\label{Diagram6}
\end{figure}

The purpose of this section is to evaluate the first class of the extra 
diagrams that are needed to cancel the gauge dependence of the scalar 
self-mass on de Sitter background. These diagrams are depicted in 
Figure~\ref{Diagram6}. 
\noindent We begin by evaluating the primitive diagram on flat space 
background. Although this is not needed to cancel gauge dependence, the
exercise is useful because the diagram's basic reduction and divergence
structure are identical to the de Sitter case, which is evaluated next.
The section closes with a discussion of renormalization.

\subsection{Primitive Result on Flat Space}

The leftmost diagram of Figure~\ref{Diagram6} represents the correction
to leg \#2. Its analytic form is,
\begin{eqnarray}
\lefteqn{V^{\rm flat}_{6,2}(x;x') = \!\! \int \!\! d^Dy \, U_2^*(y)
\!\times\! i\kappa \Bigl[ \overleftarrow{\partial}^{\mu}_{y}  
\overrightarrow{\partial}_{y}^{\nu} \!-\! \tfrac12 \eta^{\mu\nu} 
(\overleftarrow{\partial}_{y} \!\!\cdot\!\! \overrightarrow{\partial}_{y} 
\!+\! m^2) \Bigr] } \nonumber \\
& & \hspace{1.9cm} \times i\Delta^{\rm flat}_{m}(y;x) \!\times\! 
(-\tfrac{i}{2} \kappa \lambda \eta^{\rho\sigma} ) i [\mbox{}_{\mu\nu} 
\Delta^{\rm flat}_{\rho\sigma}](y;x) \!\times\! i\Delta(x;x') (-i\lambda) 
\; , \qquad \label{Vflat62}
\end{eqnarray}
where the flat space mode function is
\begin{equation}
\lim_{H \rightarrow 0} u_2^*(y) \equiv U_2^*(y) = \tfrac{\exp[i\omega_2 
y^0 - i \vec{k}_2 \cdot \vec{y}]}{\sqrt{2 \omega_2}} \qquad , \qquad  
\omega_2 \equiv \sqrt{m^2 + \Vert \vec{k}_2 \Vert^2} \; . \label{flatmode}
\end{equation}
In the flat 
space limit the massless, minimally coupled scalar propagator (\ref{DeltaA}) 
and the graviton propagator (\ref{gravprop}) become,
\begin{eqnarray}
i \Delta(x;y) &\!\!\! = \!\!\!& \tfrac{\Gamma(\frac{D}2 - 1)}{4 
\pi^{\frac{D}2}} \tfrac1{\Delta x^{D-2}} \; , \qquad \label{flatscalar} \\
i [\mbox{}_{\mu\nu} \Delta^{\rm flat}_{\rho\sigma}](y;x) &\!\!\! = \!\!\!&
[2 \eta_{\mu (\rho} \eta_{\sigma) \nu} - \tfrac{2}{D-2} \eta_{\mu\nu} 
\eta_{\rho\sigma}] i \Delta(x;y) \; . \qquad \label{flatgrav}
\end{eqnarray}
The flat space limit of the massive scalar propagator (\ref{DeltaM}) has
the expansion,
\begin{eqnarray}
\lefteqn{i\Delta^{\rm flat}_{m}(y;x) = \tfrac{\Gamma(\frac{D}2 - 1)}{4 
\pi^{\frac{D}2}} \tfrac1{\Delta x^{D-2}} - \tfrac{\Gamma(\frac{D}2 -1)}{
16 (\frac{D}2 - 2) \pi^{\frac{D}2}} \tfrac{m^2}{\Delta x^{D-4}} +
\tfrac{\Gamma(1 - \frac{D}2)}{(4 \pi)^{\frac{D}2}} \, m^{D-2} } 
\nonumber \\
& & \hspace{-0.5cm} + \tfrac{m^2}{16 \pi^2} \sum_{n=1}^{\infty}
\tfrac{(\frac14 m^2 \Delta x^2)^n}{n! (n+1)!} \Bigl\{ \ln(\tfrac14 m^2 
\Delta x^2) \!-\! \psi(n\!+\!2) \!-\! \psi(n\!+\!1) \Bigr\} + O(D\!-\!4) \; .
\qquad \label{flatmassive}
\end{eqnarray}
Because the diagram is linearly divergent, we need only the first term.

Substituting (\ref{flatgrav}) allows us to write,
\begin{equation}
V^{\rm flat}_{6,2}(x;x') = -i \kappa^2 \lambda^2 \!\! \int \!\! d^Dy \, 
U_2^*(y) [\overleftarrow{\partial}_{y} \!\!\cdot\!\! 
\overrightarrow{\partial}_{y} \!+\! \tfrac{D m^2}{D-2} ] 
i\Delta^{\rm flat}_{m}(y;x) \!\times\! i\Delta(y;x) i\Delta(x;x') \; . 
\label{Vflat62step2} 
\end{equation}
Further progress comes from isolating the part of the product of the 
two scalar propagators for which the integral over $y^{\mu}$ diverges
in $D=4$ spacetime dimensions,
\begin{eqnarray}
\lefteqn{i \Delta^{\rm flat}_{m}(y;x) i\Delta(y;x) = 
\tfrac{\Gamma^2(\frac{D}2 -1)}{16 \pi^D} \tfrac1{\Delta x^{2D-4}}
} \nonumber \\
& & \hspace{0cm} + \tfrac{m^2}{64 \pi^4 \Delta x^2} \sum_{n=0}^{\infty} 
\tfrac{(\frac14 m^2 \Delta x^2)^n}{n! (n+1)!} \Bigl\{ \ln(\tfrac14 m^2
\Delta x^2) \!-\! \psi(n\!+\!2) \!-\! \psi(n\!+\!1) \Bigr\} + 
\dots \; . \qquad \label{scalarproduct}
\end{eqnarray}
Only the term on the first line of expression (\ref{scalarproduct}) 
fails to be integrable with respect to $y^{\mu}$ for $D=4$; the terms 
on the second line are integrable, and the neglected terms ($\dots$) 
are not only integrable but actually zero in $D=4$ dimensions. To 
analyze the divergent term we first extract a d'Alembertian and then 
add zero in the form of the propagator equation for $i\Delta(y;x)$ 
\cite{Onemli:2002hr},
\begin{eqnarray}
\tfrac1{\Delta x^{2D-4}} &\!\!\! = \!\!\!& \tfrac{\partial_{y}^2}{2 (D-3) 
(D-4)} \tfrac1{\Delta x^{2D-6}} \; , \\
&\!\!\! = \!\!\!& \tfrac{\partial_{y}^2}{2 (D-3) (D-4)} \Bigl[ 
\tfrac1{\Delta x^{2D-6}} - \tfrac{\mu^{D-4}}{\Delta x^{D-2}} \Bigr] +
\tfrac{\mu^{D-4}}{2 (D-3) (D-4)} \tfrac{4 \pi^{\frac{D}2} i 
\delta^D(y - x)}{\Gamma(\frac{D}2 - 1)} \; . \qquad 
\end{eqnarray}
The factor of $\partial_{y}^2$ can be partially integrated to act on
the mode function, and the term inside the square brackets is both
integrable and of order $(D-4)$, so we can write,
\begin{equation}
\tfrac1{\Delta x^{2D-4}} = \tfrac{\mu^{D-4}}{2 (D-3) (D-4)} 
\tfrac{4 \pi^{\frac{D}2} i \delta^D(y - x)}{\Gamma(\frac{D}2 - 1)}
- \tfrac14 \partial_{y}^2 [\tfrac{\ln(\mu^2 \Delta x^2)}{\Delta x^2}]
+ \dots \; . \label{localdiv}
\end{equation}
Hence we can write,
\begin{eqnarray}
i\Delta_m^{\rm flat}(y;x) \!\times\! i\Delta(y;x) &\!\!\! = \!\!\!& 
\mathcal{K} i \delta^D(y \!-\! x) + {\rm Finite} \; , \label{product1} \\
\partial^y_{\mu} i\Delta_m^{\rm flat}(y;x) \!\times\! i\Delta(y;x) 
&\!\!\! = \!\!\!& \tfrac12 \mathcal{K} \partial^y_{\mu} 
i\delta^D(y \!-\! x) + {\rm Finite} \; , \label{product2} 
\end{eqnarray}
where the divergent constant $\mathcal{K}$ is, 
\begin{equation}
\mathcal{K} \equiv \tfrac{\Gamma(\frac{D}2 -1)}{4 \pi^{\frac{D}2}} 
\tfrac{\mu^{D-4}}{2 (D-3) (D-4)} \; . \label{Kdef}
\end{equation}

Substituting relations (\ref{product1}-\ref{product2}) into 
(\ref{Vflat62step2}), and exploiting the wave function equation
$\partial^2_y U_2^*(y) = m^2 U^*_2(y)$, gives the final primitive result,
\begin{equation}
V^{\rm flat}_{6,2}(x;x') = \tfrac{\kappa^2 m^2 \lambda^2}{2} 
(\tfrac{D+2}{D-2}) \mathcal{K} U_2^*(x) i \Delta(x;x') + {\rm Finite} 
\; . \label{Vflat62step5}
\end{equation}
Although it would have been possible to write down an explicit expression
for the nonlocal, finite part, no purpose would be served by this. We will
omit similar finite contributions for other corrections.

\subsection{Primitive Result on de Sitter}

The first steps in reducing the de Sitter generalization of expression 
(\ref{Vflat62}) are the same as in flat space,
\begin{eqnarray}
\lefteqn{V_{6,2}(x;x') = \!\! \int \!\! d^Dy \, u_2^*(y) \!\times\! 
i\kappa \Bigl[ \overleftarrow{\partial}^{\mu}_{y} a^{D-2}_{y} 
\overrightarrow{\partial}_{y}^{\nu} \!-\! \tfrac12 \eta^{\mu\nu} 
(\overleftarrow{\partial}_{y} a^{D-2}_{y} \!\cdot\! \overrightarrow{
\partial}_{y} \!+\! a_y^D m^2) \Bigr] } \nonumber \\
& & \hspace{0.9cm} \times i\Delta_{m}(y;x) \!\times\! (-\tfrac{i}{2} 
\kappa \lambda a_x^D \eta^{\rho\sigma} ) i [\mbox{}_{\mu\nu} 
\Delta_{\rho\sigma}](y;x) \!\times\! i\Delta_A(x;x') (-i\lambda a_{x'}^D) 
\; , \qquad \label{V62} \\
& & \hspace{-0.5cm} = -\tfrac{i}{2} \kappa^2 \lambda^2 \!\! \int \!\! 
d^Dy \, u_2^*(y) \Bigl\{ \overleftarrow{\partial}^{\mu}_{y} a^{D-2}_{y} 
\overrightarrow{\partial}_{y}^{\nu} \!-\! \tfrac12 \eta^{\mu\nu} 
(\overleftarrow{\partial}_{y} a^{D-2}_{y} \!\cdot\! \overrightarrow{
\partial}_{y} \!+\! a_y^D m^2) \Bigr\} i\Delta_{m}(y;x) \nonumber \\
& & \hspace{1.5cm} \times \Bigl[-\tfrac{4 \eta_{\mu\nu}}{D-2} i\Delta_C(y;x)
-\tfrac{4 \overline{\eta}_{\mu\nu}}{D-3} i\Delta_{AC}(y;x)\Bigr] 
\!\times\! (a_x a_x')^D i\Delta_A(x;x') \; , \qquad \label{V61step2} \\
& & \hspace{-0.5cm} = -i \kappa^2 \lambda^2 \!\! \int \!\! d^Dy \, 
u_2^*(y) \Bigl\{ [\overleftarrow{\partial}_{y} a^{D-2}_{y} \!\!\cdot\!\! 
\overrightarrow{\partial}_{y} i\Delta_m(y;x) \!+\! \tfrac{D m^2 a^D_y}{D-2} 
i\Delta_m(y;x) ] i\Delta_C(y;x) \nonumber \\
& & \hspace{0cm} - \tfrac{2}{D-3} \overleftarrow{\nabla}_{y} a^{D-2}_{y} 
\!\!\cdot\!\! \overrightarrow{\nabla}_{y} i\Delta_m(y;x) \!\times\! 
i\Delta_{AC}(y;x) \!+\! (\tfrac{D-1}{D-3}) [\overleftarrow{\partial}_{y} 
a^{D-2}_{y} \!\!\cdot\!\! \overrightarrow{\partial}_{y} i\Delta_m(y;x) 
\nonumber \\
& & \hspace{3.5cm} + a^D_y m^2 i\Delta_m(y;x)] i\Delta_{AC}(y;x) \!
\Bigr\} (a_x a_x')^D i\Delta_A(x;x') . \qquad \label{V61step3}
\end{eqnarray}
Because no more than a single derivative acts on $i\Delta_m \sim 
1/(y-x)^{D-2}$, and no derivatives act on $i\Delta_{AC} \sim
1/(y-x)^{D-4}$, none of the terms involving $i\Delta_{AC}$
are divergent. That leaves the terms involving products of $i\Delta_m$
and $i\Delta_C$, whose divergences can be isolated by the de Sitter
analogs of relations (\ref{product1}-\ref{product2}),
\begin{eqnarray}
i\Delta_m(y;x) \!\times\! i\Delta_C(y;x) &\!\!\! = \!\!\!& \mathcal{K}
\tfrac{i \delta^D(y - x)}{(a_y a_x)^{D-2}} + {\rm Finite} \; , \qquad
\label{dSprod1} \\
\partial^y_{\mu} i\Delta_m(y;x) \!\times\! i\Delta_C(y;x) &\!\!\! = 
\!\!\!& \tfrac{\mathcal{K}}{2} \partial^y_{\mu} \tfrac{i \delta^D(y - x)
}{(a_y a_x)^{D-2}} + {\rm Finite} \; . \qquad \label{dSprod2}
\end{eqnarray}
Using these relations and the mode function equation gives,
\begin{equation}
V_{6,2}(x;x') = \tfrac{\kappa^2 m^2 \lambda^2}{2} (\tfrac{D+2}{D-2}) 
\mathcal{K} u^*_2(x) a_x^4 a^D_{x'} i\Delta_A(x;x') + {\rm Finite} 
\; . \label{V62step4}
\end{equation}

\subsection{Renormalization}

The divergences of Diagram 6 can be removed by a renormalization of the
coupling constant $\lambda$ in vertex (\ref{phi2psi}),
\begin{equation}
\Delta \mathcal{L} = -\tfrac12 c_1 \phi \psi^2 \sqrt{-g} \; . \label{DeltaL6}
\end{equation}
An insertion of this vertex at $x^{\mu}$ will absorb the divergences from
legs \# 1 and \# 2 if the coefficient $c_1$ is,
\begin{equation}
c_1 = \kappa^2 m^2 \lambda (\tfrac{D+2}{D-2}) \mathcal{K} \; . \label{c1def}
\end{equation}
When we include the corrections from all four legs (without the external 
wave functions), the secular part of the renormalized result is,
\begin{equation}
V_{6}(x;x') = -\tfrac{3 \kappa^2 m^2 \lambda^2}{8 \pi^2} \, \ln(a_x a_{x'}) 
\!\times\! (a_x a_{x'})^4 i\Delta_{A}(x;x') \; . \label{V6total}
\end{equation}
Inserting propagators, as the Introduction has described, allows us to 
recognize the contribution to the self-mass from Diagram 6,
\begin{equation}
-i M^2_{6}(x;x') = - \tfrac{3 \kappa^2 m^2}{8 \pi^2} \Bigl\{\mathcal{D}_{x}
[\ln(a_x) i\delta^4(x \!-\! x')]  \!+\! \mathcal{D}_{x'} [\ln(a_{x'}) i 
\delta^4(x \!-\! x')] \Bigr\} \; . \label{Msq6total}
\end{equation}

\section{Diagram 7}

\begin{figure}[ht]
\centering
\includegraphics[width=7cm]{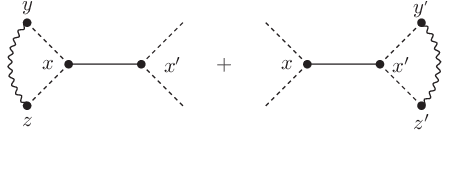}
\vskip -1cm
\caption{\footnotesize Diagram 7 correction. Solid lines
correspond to the massless scalar, dashed lines correspond to the
massive scalar, and wavy lines are gravitons.}
\label{Diagram7}
\end{figure}

The purpose of this section is to evaluate the second class of the extra 
diagrams that are needed to cancel the gauge dependence of the scalar 
self-mass on de Sitter background. These diagrams are depicted in 
Figure~\ref{Diagram7}. 
\noindent As in the previous section, we begin by evaluating the primitive 
diagram on flat space background, then derive the de Sitter result. The 
section closes with a discussion of renormalization.

\subsection{Primitive Result on Flat Space}

The lefthand diagram of Figure~\ref{Diagram7} gives corrections to legs \#1
and \#2. Its analytic form is,
\begin{eqnarray}
\lefteqn{ V^{\rm flat}_{7,12}(x;x') = \!\! \int \!\! d^Dy \, U_2^*(y)
\!\times\! i\kappa \Bigl[ \overleftarrow{\partial}^{\mu}_{y}  
\overrightarrow{\partial}_{y}^{\nu} \!-\! \tfrac12 \eta^{\mu\nu} 
(\overleftarrow{\partial}_{y} \!\!\cdot\!\! \overrightarrow{\partial}_{y} 
\!+\! m^2) \Bigr] i \Delta^{\rm flat}_{m}(y;x) } \nonumber \\
& & \hspace{0.5cm} \times\!\! \int \!\! d^Dz \, U_1(z) \!\times\! i \kappa
\Bigl[ \overleftarrow{\partial}^{\rho}_{z} \overrightarrow{\partial}_{z}^{\sigma} 
\!-\! \tfrac12 \eta^{\rho\sigma} (\overleftarrow{\partial}_{z} \!\!\cdot\!\! 
\overrightarrow{\partial}_{z} \!+\! m^2) \Bigr] i \Delta^{\rm flat}_{m}(z;x)
\nonumber \\
& & \hspace{5cm} \times i [\mbox{}_{\mu\nu} \Delta^{\rm flat}_{\rho
\sigma}](y;z) \!\times\! -i\lambda \!\times\! i\Delta(x;x') \!\times\! 
-i\lambda \; . \qquad \label{Vflat712} 
\end{eqnarray}
Recall that the flat space mode functions and propagators appear in 
expressions (\ref{flatmode}-\ref{flatmassive}). Contacting the tensor indices 
gives,
\begin{eqnarray}
\lefteqn{ V^{\rm flat}_{7,12}(x;x') = \kappa^2 \lambda^2 \!\! \int \!\! d^Dy
\, U_2^*(y) \!\! \int \!\! d^Dz \, U_1(z) \Bigl\{ \overleftarrow{\partial}_y
\!\! \cdot\!\! \overleftarrow{\partial}_{z} \overrightarrow{\partial}_{y}
\!\! \cdot \!\! \overrightarrow{\partial}_{z} \!+\! \overleftarrow{\partial}_y
\!\! \cdot\!\! \overrightarrow{\partial}_{z} \overrightarrow{\partial}_{y}
\!\! \cdot \!\! \overleftarrow{\partial}_{z} } \nonumber \\
& & \hspace{-0.5cm} - \overleftarrow{\partial}_y \!\! \cdot\!\! 
\overrightarrow{\partial}_{y} \overleftarrow{\partial}_z \!\! \cdot\!\! 
\overrightarrow{\partial}_{z} \!-\! m^2 ( \overleftarrow{\partial}_y \!\! 
\cdot\!\! \overrightarrow{\partial}_{y} \!+\! \overleftarrow{\partial}_z 
\!\! \cdot\!\! \overrightarrow{\partial}_{z}) \!-\! \tfrac{D m^4}{D-2}
\Bigr\} i\Delta^{\rm flat}_m(y;x) i\Delta^{\rm flat}_m(z;x) \nonumber \\
& & \hspace{8.5cm} \times i\Delta(y;z) i\Delta(x;x') \; . \qquad 
\label{Vflat712step2}
\end{eqnarray} 

Because Diagram 7 contains three internal propagators, it can only 
contribute a divergence when two derivatives act on the internal
propagators. Hence we just need the first three terms within the curly 
brackets of (\ref{Vflat712step2}),
\begin{equation}
\overleftarrow{\partial}_y^{\mu} \overleftarrow{\partial}_z^{\nu} 
[\eta_{\mu\nu} \partial_y \!\cdot\! \partial_z \!+\! \partial^y_{\nu}
\partial^z_{\mu} \!-\! \partial^y_{\mu} \partial^z_{\nu} ] 
i\Delta_m^{\rm flat}(y;x) i\Delta_m^{\rm flat}(z;x) \!\times\! i\Delta(y;z)
\; . \label{threedees}
\end{equation}
The first of these three terms is simplest. Because the propagators depend
only on the differences of two coordinates we can convert them to derivatives
with respect to $x^{\mu}$ and then extract them,
\begin{eqnarray}
\lefteqn{ \partial^{\mu}_{y} i\Delta^{\rm flat}_{m}(y;x) \!\times\!
\partial^z_{\mu} i\Delta^{\rm flat}_{m}(z;x) = \partial_{x}^{\mu} 
i\Delta^{\rm flat}_{m}(y;x) \!\times\! \partial^{x}_{\mu} 
i\Delta^{\rm flat}_{m}(z;x) \; , } \\
& & \hspace{-0.5cm} = \tfrac12 \partial_{x}^2 [i\Delta^{\rm flat}_{m}(y;x)
i\Delta^{\rm flat}_{m}(z;x)] \nonumber \\
& & \hspace{2.2cm} - \tfrac12 \partial_x^2 i\Delta^{\rm flat}_{m}(y;x) 
i\Delta^{\rm flat}_{m}(z;x) \!-\! \tfrac12 i\Delta^{\rm flat}_{m}(y;x) 
\partial_x^2 i\Delta^{\rm flat}_{m}(z;x) \; , \qquad \\
& & \hspace{-0.5cm} = \tfrac12 (\partial_{x}^2 \!-\! 2 m^2) 
[i\Delta^{\rm flat}_{m}(y;x) i\Delta^{\rm flat}_{m}(z;x)] \!-\! \tfrac{i}{2}
[{\scriptstyle \delta^D(y - x)} \!+\! {\scriptstyle \delta^D(z - x)}] 
i\Delta^{\rm flat}_{m}(y;z) \; . \qquad \label{factorB}
\end{eqnarray}
When multiplied by the massless propagator $i\Delta(y;z)$, the first term
in expression (\ref{threedees}) is ultraviolet finite. The divergence arises
from the product (\ref{product1}), which we repeat here,
\begin{equation}
i\Delta^{\rm flat}_{m}(y;z) i\Delta(y;z) = \mathcal{K} \, i\delta^D(y \!-\! z) 
+ {\rm Finite} \; . \label{FactorBdiv}
\end{equation}
Hence we conclude that the divergent contribution from the first term inside
the curly brackets of expression (\ref{Vflat712step2}) is,
\begin{equation}
\kappa^2 \lambda^2 \mathcal{K} \, \partial^{\mu} U_2^*(x) \!\times\! 
\partial_{\mu} U_1(x) \!\times\! i\Delta(x;x') \; . \label{term1div}
\end{equation}

The action of uncontracted derivatives on the two massive propagators is
complicated, but its divergent part is proportional to the trace,
\begin{equation}
\partial^{\mu}_{y} i\Delta^{\rm flat}_{m}(y;x) \!\times\! \partial^{\nu}_{z}
i\Delta^{\rm flat}_{m}(z;x) = \tfrac1{D} \eta^{\mu\nu} \partial^{\alpha}_{y} 
i\Delta^{\rm flat}_{m}(y;x) \!\times\! \partial^{z}_{\alpha} 
i\Delta^{\rm flat}_{m}(z;x) + {\rm Finite} \; . \label{term2prime}
\end{equation}
It follows that the divergent contributions from the second and third terms
in (\ref{threedees}) cancel. Hence the left hand diagram of Figure~\ref{Diagram7}
is,
\begin{equation}
V^{\rm flat}_{7,12}(x;x') = \kappa^2 \lambda^2 \mathcal{K} \, 
\partial^{\mu} U_2^*(x) \!\times\! \partial_{\mu} U_1(x) \!\times\! 
i\Delta(x;x') + {\rm Finite} \; . \label{Vflat712step4}
\end{equation}

\subsection{Primitive Result on de Sitter}

The reduction of Diagram 7 on de Sitter begins the same as in flat space,
\begin{eqnarray}
\lefteqn{ V_{7,12} = \!\! \int \!\! d^Dy \, u_2^*(y)
\!\times\! i\kappa \Bigl[ \overleftarrow{\partial}^{\mu}_{y} a_y^{D-2}  
\overrightarrow{\partial}_{y}^{\nu} \!-\! \tfrac12 \eta^{\mu\nu} 
(\overleftarrow{\partial}_{y} a_y^{D-2} \!\!\cdot\!\! 
\overrightarrow{\partial}_{y} \!+\! a_y^D m^2) \Bigr] i \Delta_{m}(y;x) } 
\nonumber \\
& & \hspace{0cm} \times\!\! \int \!\! d^Dz \, u_1(z) \!\times\! i \kappa
\Bigl[ \overleftarrow{\partial}^{\rho}_{z} a_z^{D-2} 
\overrightarrow{\partial}_{z}^{\sigma} \!-\! \tfrac12 \eta^{\rho\sigma} 
(\overleftarrow{\partial}_{z} a_z^{D-2} \!\!\cdot\!\! 
\overrightarrow{\partial}_{z} \!+\! a_z^D m^2) \Bigr] i \Delta_{m}(z;x)
\nonumber \\
& & \hspace{3.5cm} \times i [\mbox{}_{\mu\nu} \Delta_{\rho\sigma}](y;z) 
\!\times\! -i\lambda a_x^D \!\times\! i\Delta_A(x;x') \!\times\! 
-i\lambda a_{x'}^D \; , \qquad \label{V712} \\
& & \hspace{-0.5cm} = \kappa^2 \lambda^2 \!\! \int \!\! d^Dy \, u_2^*(y) 
\Bigl[ \overleftarrow{\partial}^{\mu}_{y} a_y^{D-2}  
\overrightarrow{\partial}_{y}^{\nu} \!-\! \tfrac12 \eta^{\mu\nu}
(\overleftarrow{\partial}_{y} a_y^{D-2} \!\!\cdot\!\! 
\overrightarrow{\partial}_{y} \!+\! a_y^D m^2) \Bigr] i\Delta_m(y;x) 
\qquad \nonumber \\
& & \hspace{0.5cm} \times\!\! \int \!\! d^Dz \, u_1(z) \Bigl[ 
\overleftarrow{\partial}^{\rho}_{z} a_z^{D-2} 
\overrightarrow{\partial}_{z}^{\sigma} \!-\! \tfrac12 \eta^{\rho\sigma}
(\overleftarrow{\partial}_{z} a_z^{D-2} \!\!\cdot\!\! 
\overrightarrow{\partial}_{z} \!+\! a_z^D m^2) \Bigr] i \Delta_{m}(z;x)
\nonumber \\
& & \hspace{1cm} \times \Bigl\{ [2\eta_{\mu (\rho} \eta_{\sigma)\nu} 
\!-\! \tfrac2{D-2} \eta_{\mu\nu} \eta_{\rho\sigma}] i\Delta_C(y;z) \!+\!
[\mbox{}_{\mu\nu} T^A_{\rho\sigma}] i\Delta_{AC}(y;z) \nonumber \\
& & \hspace{4.5cm} + [\mbox{}_{\mu\nu} T^B_{\rho\sigma}] i\Delta_{BC}(y;z) 
\Bigr\} \!\times\! (a_x a_{x'})^D i\Delta_A(x;x') \; . \qquad 
\label{V712step2}
\end{eqnarray}
Isolating the divergences of (\ref{V712step2}) is almost identical to
that of its flat space cousin (\ref{Vflat712step2}) because there are
three internal propagators, and at most two derivatives can act on them.
This means we can drop the less singular parts of the graviton propagator
proportional to $i\Delta_{AC}(y;z)$ and $i\Delta_{BC}(y;z)$. We can also
drop terms with fewer than two internal derivatives,
\begin{eqnarray}
\lefteqn{ V_{7,12}(x;x') = \kappa^2 \lambda^2 \!\! \int \!\! d^Dy \, 
\partial_y^{\mu} u_2^*(y) \!\! \int \!\! d^Dz \, \partial_z^{\nu} u_1(z) 
\!\times\! (a_y a_z)^{D-2} \Bigl\{\eta_{\mu\nu} \partial^y_{\alpha}
i \Delta_m(y;x) } \nonumber \\
& & \hspace{0cm} \times \partial_z^{\alpha} i\Delta_m(z;x) \!+\! 
\partial^{y}_{\nu} i\Delta_m(y;x) \partial^z_{\mu} i\Delta_m(z;x) \!-\! 
\partial^y_{\mu} i\Delta_m(y;x) \partial^z_{\nu} i\Delta_m(z;x) \Bigr\} 
\nonumber \\
& & \hspace{5.2cm} \times i\Delta_C(y;z) (a_x a_{x'})^D i\Delta_A(x;x') \!+\!
{\rm Finite} \; . \qquad \label{V712step3}
\end{eqnarray}

Because the three terms in expression (\ref{V712step3}) are only 
logarithmically divergent, we need just the first terms in the expansions
of $i\Delta_m$ and $i\Delta_C$. Except for some scale factors, this first
term is the same as in flat space, and we can forget about cases where 
derivatives act on scale factors. All of this means that the divergent 
part of $V_{7,12}$ can be read off from its flat space cousin 
(\ref{Vflat712step4}),
\begin{equation}
V_{7,12}(x;x') = \kappa^2 \lambda^2 \mathcal{K} a_x^2 \partial^{\mu} 
u_2^*(x) \partial_{\mu} u_1(x) \!\times\! i\Delta_A(x;x') a_{x'}^D 
+ {\rm Finite} \; . \qquad \label{V712step4}
\end{equation}

\subsection{Renormalization}

The counterterm needed to renormalize Diagram 7 is a sort of 
$\phi$-dependent field strength renormalization for $\psi$,
\begin{equation}
\Delta \mathcal{L} = - \tfrac12 c_2 \phi \partial_{\mu} \psi 
\partial_{\nu} \psi g^{\mu\nu} \sqrt{-g} \qquad , \qquad c_2 =
\kappa^2 \lambda \mathcal{K} \; . \label{DeltaL7}
\end{equation}
The final renormalized result for the secular part of Diagram 7 is,
\begin{eqnarray}
\lefteqn{V_7(x;x') = -\tfrac{\kappa^2 \lambda^2}{8 \pi^2} \Bigl\{ 
\ln(a_x) a_x^2 \partial^1_x \!\cdot\! \partial^2_x \!\times\! a_{x'}^4 
\!+\! a_x^4 \!\times\! \ln(a_{x'}) a_{x'}^2 \partial^3_{x'} \!\cdot\! 
\partial^4_{x'} \Bigr\} } \nonumber \\
& & \hspace{7.4cm} \times i\Delta_A(x;x') \!+\! {\rm Nonsecular} \; . 
\qquad \label{V7}
\end{eqnarray}
Here the superscripts on the derivatives indicate the leg on which they
act. 

To infer the self-mass from (\ref{V7}) we first exploit momentum 
conservation (\ref{conservation}) in $D=4$ to re-express the derivatives 
on external legs,
\begin{equation}
a_x^2 \partial^1_x \!\cdot\! \partial^2_x = \tfrac12 
\widetilde{\mathcal{D}}_x - \tfrac12 \widetilde{\mathcal{D}}^1_{x} - 
\tfrac12 \widetilde{\mathcal{D}}^2_{x} = \tfrac12 
\widetilde{\mathcal{D}}_x - m^2 a_x^4 \; . \label{V7reduce1}
\end{equation}
A single secular contribution comes from $\widetilde{\mathcal{D}}_x$ 
acting on $\ln(a_x) i\Delta_A(x;x')$,
\begin{equation}
\tfrac12 \widetilde{\mathcal{D}}_x [\ln(a_x) i\Delta_A(x;x')] = 
\tfrac12 \ln(a_x) i\delta^4(x \!-\! x') + {\rm Nonsecular} \; .
\label{V7reduce2}
\end{equation} 
We can therefore write the secularly enhanced part of Diagram 7 as,
\begin{eqnarray}
\lefteqn{V_7(x;x') \longrightarrow (-i\lambda a_x^4) \Bigl\{ -
\tfrac{\kappa^2 m^2}{8\pi^2} \ln(a_x a_{x'}) i\Delta_A(x;x') \!+\! 
\tfrac{\kappa^2}{8\pi^2} \tfrac{\ln(a_x) i\delta^4(x - x')}{a_x^4} \Bigr\} 
(-i \lambda a_{x'}^4) } \nonumber \\
& & \hspace{-0.5cm} = (-i \lambda a_x^4) \!\! \int \!\! d^4z \, i\Delta_A(x;z)
\!\! \int \!\! d^4z' \, i\Delta_A(x';z') \Bigl\{ \tfrac{\kappa^2 m^2}{8\pi^2}
\mathcal{D}_z [\ln(a_z) i\delta^4(z \!-\! z')] \qquad \nonumber \\
& & \hspace{0.8cm} + \tfrac{\kappa^2 m^2}{8\pi^2} \mathcal{D}_{z'} [\ln(a_{z'}) 
i\delta^4(z \!-\! z')] - \tfrac{\kappa^2}{8\pi^2} \mathcal{D}_z 
\mathcal{D}_{z'} [\tfrac{\ln(a_z) i\delta^4(z - z')}{a_z^4} ] \Bigr\} 
(-i \lambda a_{x'}^4) \; . \qquad \label{V7final}
\end{eqnarray}
It follows that the secular contribution to the self-mass from
Diagram 7 is,
\begin{eqnarray}
\lefteqn{ -i M^2_{7}(x;x') = \tfrac{\kappa^2 m^2}{8\pi^2} \Bigl\{ 
\mathcal{D}_x [\ln(a_x) i\delta^4(x \!-\! x')] \!+\! \mathcal{D}_{x'} 
[\ln(a_{x'}) i\delta^4(x \!-\! x')] \Bigr\} } \nonumber \\
& & \hspace{7.5cm} - \tfrac{\kappa^2}{8\pi^2} \mathcal{D}_x 
\mathcal{D}_{x'} [\tfrac{\ln(a_x) i\delta^4(x - x')}{a_x^4} ] \; .
\qquad \label{M7}
\end{eqnarray}

\section{Wavefunction Corrections}

\begin{figure}[ht]
\centering
\includegraphics[width=7cm]{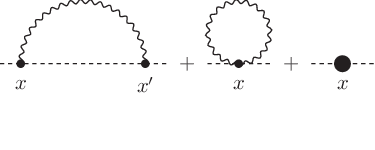} 
\vskip -1.5cm
\caption{\footnotesize 1-loop corrections to the massive scalar
self-mass $-i M^2(x;x')$. Dashed lines correspond to the massive 
scalar, and wavy lines are gravitons. The two leftmost diagrams
give the primitive contribution and the righthand diagram represents
the counterterm.}
\label{Massive}
\end{figure}

The purpose of this section is to derive the 1-loop graviton correction
to the massive scalar self-mass, and then use it to compute the field 
strength renormalization which corrects external legs. The diagrammatic
representation for the self-mass is depicted in Figure~\ref{Massive}.
As in the previous two sections, we first evaluate the dimensionally
regulated, primitive result for flat space and then compute the de
Sitter generalization. The next step is renormalization, following 
which we infer the secular parts of the external state corrections.

\subsection{Primitive Result for $-i M^2(x;x')$ on Flat Space}

In flat space the middle diagram of Figure~\ref{Massive} vanishes 
because it contains a coincident graviton propagator. The flat space
limit of the leftmost is,
\begin{eqnarray}
\lefteqn{ -i M^2_{\rm flat}(x;x') = i\kappa \Bigl[ 
\overleftarrow{\partial}^{\mu} \overrightarrow{\partial}^{\nu} 
\!\!-\! \tfrac12 \eta^{\mu\nu} (\overleftarrow{\partial} \!\!\cdot\!\! 
\overrightarrow{\partial} \!+\! m^2) \Bigr] i \Delta^{\rm flat}_{m}(x;x') 
} \nonumber \\
& & \hspace{3cm} \times i \kappa \Bigl[ {\overleftarrow{\partial}'}^{\rho}
{\overrightarrow{\partial}'}^{\sigma} \!\!\!-\! \tfrac12 \eta^{\rho\sigma}
({\overleftarrow{\partial}'} \!\!\cdot\!\! {\overrightarrow{\partial}'} 
\!+\! m^2) \Bigr] \!\times\! i [\mbox{}_{\mu\nu} \Delta^{\rm flat}_{
\rho\sigma}](x;x') \; , \qquad \label{Mflat}
\end{eqnarray}
\begin{eqnarray}
& & \hspace{-0.5cm} = -\kappa^2 \Bigl\{\overleftarrow{\partial} 
\!\!\cdot\!\! {\overleftarrow{\partial}'} \overrightarrow{\partial} 
\!\!\cdot\!\! {\overrightarrow{\partial}'} \!+\! \overleftarrow{\partial} 
\!\!\cdot\!\! {\overrightarrow{\partial}'} \overrightarrow{\partial} 
\!\!\cdot\!\! {\overleftarrow{\partial}'} \!-\! \overleftarrow{\partial} 
\!\!\cdot\!\! \overrightarrow{\partial} {\overleftarrow{\partial}'} 
\!\!\cdot\!\! {\overrightarrow{\partial}'} \!-\! m^2 
(\overleftarrow{\partial} \!\!\cdot\!\! \overrightarrow{\partial} \!+\!
{\overleftarrow{\partial}'} \!\!\cdot\!\! {\overrightarrow{\partial}'})
\nonumber \\
& & \hspace{6.5cm} - \tfrac{D m^4}{D-2} \Bigr\} i\Delta^{\rm flat}_{m}(x;x')
\!\times\! i\Delta(x;x') \; . \qquad \label{Mflatstep2}
\end{eqnarray}
We remind the reader that the various flat space propagators are defined
in expressions (\ref{flatscalar}-\ref{flatmassive}). Because $\partial_{\mu}
i\Delta_m(x;x') = -\partial'_{\mu} i\Delta_m(x;x')$ the 1st and 3rd terms
of (\ref{Mflatstep2}) cancel,
\begin{eqnarray}
\lefteqn{-i M^2_{\rm flat}(x;x') = -\kappa^2 \Bigl\{\overleftarrow{\partial} 
\!\!\cdot\!\! {\overrightarrow{\partial}'} \overrightarrow{\partial} 
\!\!\cdot\!\! {\overleftarrow{\partial}'} \!-\! m^2 
(\overleftarrow{\partial} \!\!\cdot\!\! \overrightarrow{\partial} \!+\!
{\overleftarrow{\partial}'} \!\!\cdot\!\! {\overrightarrow{\partial}'}) }
\nonumber \\
& & \hspace{6.5cm} - \tfrac{D m^4}{D-2} \Bigr\} i\Delta^{\rm flat}_{m}(x;x') 
\!\times\! i\Delta(x;x') \; . \qquad \label{Mflatstep3}
\end{eqnarray}

At this point we digress to consider some relations involving functions of
the Minkowski interval $\Delta x^2 = (x-x')^2$. A function of $\Delta x^2$
obeys,
\begin{eqnarray}
\partial_{\mu} f(\Delta x^2) = 2 \Delta x_{\mu} f'(\Delta x^2) \qquad , 
\qquad \partial'_{\nu} f(\Delta x^2) = -2 \Delta x_{\nu} f'(\Delta x^2) 
\; , \qquad \label{df} \\
\partial_{\mu} \partial'_{\nu} f(\Delta x^2) = -2\eta_{\mu\nu} f'(\Delta x^2)
- 4 \Delta x_{\mu} \Delta x_{\nu} f''(\Delta x^2) \; . \qquad \label{ddf}
\end{eqnarray}
Now note from expression (\ref{flatscalar}) that $i\Delta(x;x')$ is 
proportional to $1/\Delta x^{D-2}$. This suggests that we consider products
of $1/\Delta x^{D-2}$ times derivatives of $f(\Delta x^2)$,
\begin{equation}
\tfrac{\partial_{\mu} f(\Delta x^2)}{\Delta x^{D-2}} = \partial_{\mu}
I[\tfrac{f'(\Delta x^2)}{\Delta x^{D-2}}] \qquad , \qquad 
\tfrac{\partial'_{\nu} f(\Delta x^2)}{\Delta x^{D-2}} = \partial'_{\nu} 
I[\tfrac{f'(\Delta x^2)}{\Delta x^{D-2}}] \; . \label{onederiv}
\end{equation}
Here and henceforth, the symbol $I[f]$ denotes the indefinite integral of
$f(\Delta x^2)$ with respect to $\Delta x^2$. The corresponding result for
two derivatives is,
\begin{equation}
\tfrac{\partial_{\mu} \partial'_{\nu} f(\Delta x^2)}{\Delta x^{D-2}} =
\partial_{\mu} \partial'_{\nu} I^2[\tfrac{f''(\Delta x^2)}{\Delta x^{D-2}}]
+ ({\scriptstyle D - 2}) \eta_{\mu\nu} I[\tfrac{f'(\Delta x^2)}{\Delta x^{D}}]
\; . \label{twoderiv}
\end{equation} 

The next step is to note that $i\Delta^{\rm flat}_m(x;x')$ depends only on 
$\Delta x^2$ and apply the identities (\ref{onederiv}-\ref{twoderiv}) to 
(\ref{Mflatstep3}) to infer,
\begin{eqnarray}
\lefteqn{-i M^2_{\rm flat}(x;x') = -\tfrac{\kappa^2 \Gamma(\frac{D}2 -1)}{4
\pi^{\frac{D}2}} \Bigl\{ \overleftarrow{\partial} \!\!\cdot\!\! 
{\overrightarrow{\partial}'} \Bigl[ \overrightarrow{\partial} \!\!\cdot\!\!
{\overleftarrow{\partial}'} I^2[\tfrac{i {\Delta^{\rm flat}_m}''}{\Delta x^{D-2}}] 
\!+\! {\scriptstyle D (D-2)} I[\tfrac{i{\Delta^{\rm flat}_m}'}{\Delta x^{D}}] 
\Bigr] } \nonumber \\
& & \hspace{5cm} + m^2 (\overleftarrow{\partial}^2 \!+\! 
{\overrightarrow{\partial}'}^2 ) I[\tfrac{i {\Delta^{\rm flat}_m}'}{
\Delta x^{D-2}}] \!-\! \tfrac{D m^4}{D-2} \tfrac{i\Delta^{\rm flat}_m}{
\Delta x^{D-2}} \Bigr\} \; . \qquad \label{Mflatstep4}
\end{eqnarray}
The potentially divergent contributions are,
\begin{eqnarray}
I^2[\tfrac{i{\Delta^{\rm flat}_m}''}{\Delta x^{D-2}}] &\!\!\! = \!\!\!& 
\tfrac{\Gamma(\frac{D}2+1)}{4 \pi^{\frac{D}2}} \tfrac1{(D-2) (D-1) 
\Delta x^{2D-4}} + \dots \; , \label{div1} \\
I[\tfrac{i {\Delta^{\rm flat}_m}'}{\Delta x^D}] &\!\!\! = \!\!\!& 
\tfrac{\Gamma(\frac{D}2)}{4 \pi^{\frac{D}2}} \tfrac1{(D-1) \Delta x^{2D-2}} - \tfrac{\Gamma(\frac{D}2 -1)}{16 \pi^{\frac{D}2}} \tfrac{m^2}{(D-2) 
\Delta x^{2D-4}} + \dots \; , \label{div2} \qquad \\
I[\tfrac{i {\Delta^{\rm flat}_m}'}{\Delta x^{D-2}}] &\!\!\! = \!\!\!& 
\tfrac{\Gamma(\frac{D}2)}{4 \pi^{\frac{D}2}} \tfrac1{(D-2) \Delta x^{2D-4}} 
+ \dots \; , \label{div3} \\
\tfrac{i \Delta^{\rm flat}_m}{\Delta x^{D-2}} &\!\!\! = \!\!\!& 
\tfrac{\Gamma(\frac{D}2-1)}{4 \pi^{\frac{D}2}} \tfrac1{\Delta x^{2D-4}} + 
\dots \; , \label{div4} 
\end{eqnarray}
where the neglected terms ($\dots$) are both integrable and finite in $D=4$
spacetime dimensions. We isolate these divergences by extracting factors of
$\partial^2$ and then adding zero in the form of the propagator equation
for $i\Delta(x;x')$ \cite{Onemli:2002hr},
\begin{eqnarray}
\tfrac1{\Delta x^{2D-4}} &\!\!\! = \!\!\!& \tfrac{\mu^{D-4}}{2(D-3)(D-4)}
\tfrac{4 \pi^{\frac{D}2} i \delta^D(x - x')}{\Gamma(\frac{D}2 - 1)} - 
\tfrac{\partial^2}{4} [\tfrac{\ln(\mu^2 \Delta x^2)}{\Delta x^2}] + 
O({\scriptstyle D-4}) \; , \qquad \label{Dx2D4} \\
\tfrac1{\Delta x^{2D-2}} &\!\!\! = \!\!\!& \tfrac{\mu^{D-4}}{4(D-2)^2(D-3)(D-4)}
\tfrac{4 \pi^{\frac{D}2} \, \partial^2 i \delta^D(x - x')}{\Gamma(\frac{D}2 - 1)} 
- \tfrac{\partial^4}{32} [\tfrac{\ln(\mu^2 \Delta x^2)}{\Delta x^2}] + 
O({\scriptstyle D-4}) \; . \qquad \label{Dx2D2}
\end{eqnarray}
Employing relations (\ref{Dx2D4}-\ref{Dx2D2}) in expressions 
(\ref{div1}-\ref{div4}) allows us to identify the divergent part of
(\ref{Mflatstep4}),
\begin{equation}
-i M^2_{\rm flat}(x;x') = \kappa^2 m^2 \mathcal{K} \Bigl\{ -(\tfrac{D+4}4) 
\partial^2 + \tfrac{D m^2}{D-2} \Bigr\} i \delta^D(x \!-\! x') + 
{\rm Finite} \; . \qquad \label{Mflatstep5}
\end{equation}
Note that the $\partial^4$ divergences arising from expressions (\ref{div1})
and (\ref{div2}) have canceled.

\subsection{$-i M^2_{\rm 4pt}(x;x')$ on de Sitter}

Although the middle diagram of Figure~\ref{Massive} does not vanish, 
like its flat space cousin, it is still much simpler than the left hand
diagram. The analytic form is,
\begin{eqnarray}
\lefteqn{ -i M^2_{\rm 4pt}(x;x') = -\kappa^2 (\overleftarrow{\partial} 
a^{D-2} \!\!\cdot\!\! \overrightarrow{\partial} \!+\! a^D m^2) 
\, i \delta^D(x \!-\! x') } \nonumber \\
& & \hspace{3.5cm} \times \Bigl\{ \tfrac18 i [\mbox{}^{\alpha}_{~\alpha}
\Delta^{\beta}_{~\beta}](x;x) \!-\! \tfrac14 i [\mbox{}^{\alpha\beta}
\Delta_{\alpha\beta}](x;x) \Bigr\} \nonumber \\
& & \hspace{0.5cm} -\kappa^2 \overleftarrow{\partial}_{\mu}
a^{D-2} \overrightarrow{\partial}_{\nu} \, i \delta^D(x \!-\! x') \!\times\!
\Bigl\{ i [\mbox{}^{\mu \alpha} \Delta^{\nu}_{~\alpha}](x;x) \!-\! 
\tfrac12 i [\mbox{}^{\mu\nu} \Delta^{\alpha}_{~\alpha}](x;x) \Bigr\}
\; . \qquad \label{Msq4pt} 
\end{eqnarray}
The two contractions and their divergent parts are,
\begin{eqnarray}
\lefteqn{ \tfrac18 i [\mbox{}^{\alpha}_{~\alpha}
\Delta^{\beta}_{~\beta}](x;x) \!-\! \tfrac14 i [\mbox{}^{\alpha\beta}
\Delta_{\alpha\beta}](x;x) } \nonumber \\
& & \hspace{0cm} = -\tfrac14 {\scriptstyle D (D - 1)} i\Delta_{A} \!-\!
\tfrac12 {\scriptstyle (D-1)} i \Delta_B \!-\! \tfrac12 i\Delta_C = -
\tfrac{D (D-1) k \Psi_A}{4 a^{D-4}} + {\rm Finite} \; , \qquad 
\label{contract1} \\
\lefteqn{ i [\mbox{}^{\mu \alpha} \Delta^{\nu}_{~\alpha}](x;x) \!-\! 
\tfrac12 i [\mbox{}^{\mu\nu} \Delta^{\alpha}_{~\alpha}](x;x) } \nonumber \\
& & \hspace{0cm} = ({\scriptstyle D} i\Delta_A \!+\! i\Delta_B) 
\overline{\eta}^{\mu\nu} \!-\! [{\scriptstyle (D - 1)} i\Delta_B \!+\! 
2 i \Delta_C] \delta^{\mu}_{~0} \delta^{\nu}_{~0} = \tfrac{D k \Psi_A}{a^{D-4}} 
\!\times\! \overline{\eta}^{\mu\nu} \!+\! {\rm Finite} \; . \qquad 
\label{contract2}
\end{eqnarray}
It follows that the divergent contribution to the self-mass is,
\begin{eqnarray}
\lefteqn{-i M^2_{\rm 4pt}(x;x') = -\tfrac14 D (D \!-\! 1) \kappa^2 k 
\Psi_A \!\times\! [\partial^{\mu} a^2 \partial_{\mu} \!-\! a^4 m^2] 
i\delta^D(x \!-\! x') } \nonumber \\
& & \hspace{5.2cm} + D \kappa^2 k \Psi_A \!\times\! a^2 \nabla^2 
i \delta^D(x \!-\! x') + {\rm Finite} \; . \qquad \label{M4div} 
\end{eqnarray}

\subsection{$-i M^2_{\rm 3pt}(x;x')$ on de Sitter}

The de Sitter generalization of expression (\ref{Mflat}) is,
\begin{eqnarray}
\lefteqn{ -i M^2_{\rm 3pt}(x;x') = i\kappa \Bigl[ 
\overleftarrow{\partial}^{\mu} a^{D-2} \overrightarrow{\partial}^{\nu} 
\!-\! \tfrac12 \eta^{\mu\nu} (\overleftarrow{\partial} a^{D-2} 
\!\!\cdot\!\! \overrightarrow{\partial} \!+\! a^D m^2) \Bigr] 
i \Delta_{m}(x;x') } \nonumber \\
& & \hspace{0.5cm} \times i \kappa \Bigl[ {\overleftarrow{\partial}'}^{\rho}
{a'}^{D-2} {\overrightarrow{\partial}'}^{\sigma} \!\!-\! \tfrac12 
\eta^{\rho\sigma} ({\overleftarrow{\partial}'} {a'}^{D-2} \!\!\cdot\!\! 
{\overrightarrow{\partial}'} \!+\! {a'}^D m^2) \Bigr] \!\times\!
i [\mbox{}_{\mu\nu} \Delta_{\rho\sigma}](x;x') \; . \qquad \label{Msq3pt} 
\end{eqnarray}
Substituting the first line of expression (\ref{gravprop}) for the
graviton propagator gives,
\begin{eqnarray}
\lefteqn{-i M^2_{\rm 3ptC}(x;x') = -\kappa^2 \Bigl\{ (a a')^{D-2} 
\Bigl[ \overleftarrow{\partial} \!\cdot\! \overleftarrow{\partial}' 
\overrightarrow{\partial} \!\cdot\! \overrightarrow{\partial}' \!+\! 
\overleftarrow{\partial} \!\cdot\! \overrightarrow{\partial}' 
\overrightarrow{\partial} \!\cdot\! \overleftarrow{\partial}' \!-\! 
\overleftarrow{\partial} \!\cdot\! \overrightarrow{\partial} 
\overleftarrow{\partial}' \!\!\cdot\! \overrightarrow{\partial}'
\Bigr] } \nonumber \\
& & \hspace{-0.5cm} - (a a')^{D-2} m^2 ({a'}^2 \overleftarrow{\partial} 
\!\!\cdot\!\! \overrightarrow{\partial} \!+\! a^2 \overleftarrow{\partial}' 
\!\!\cdot\!\! \overrightarrow{\partial}') \!-\! \tfrac{D m^4}{D-2} (a a')^D
\Bigr\} i\Delta_m(x;x') \!\times\! i\Delta_C(x;x') . \qquad
\label{M3ptCstep2} 
\end{eqnarray}
We shall not attempt to obtain an exact expression, analogous to the
flat space result (\ref{Mflatstep4}). We instead begin by expanding the
propagators, retaining only those terms which might produce divergences,
\begin{eqnarray}
i\Delta_m(x;x') &\!\!\! = \!\!\!& \tfrac{\Gamma(\frac{D}2 - 1)}{4 
\pi^{\frac{D}2} (a a')^{\frac{D}2 -1}} \Bigl\{ \tfrac1{\Delta x^{D-2}} 
- \tfrac{[m^2 - \frac{D}2 (\frac{D}2 - 1) H^2] a a'}{2 (D - 4) 
\Delta x^{D-4}} + \dots \Bigr\} \; , \qquad \label{Dmexpansion} \\
i\Delta_C(x;x') &\!\!\! = \!\!\!& \tfrac{\Gamma(\frac{D}2 - 1)}{4 
\pi^{\frac{D}2} (a a')^{\frac{D}2 -1}} \Bigl\{ \tfrac1{\Delta x^{D-2}} 
+ \dots \Bigr\} \; . \qquad \label{DCexpansion}
\end{eqnarray}

The last line of (\ref{M3ptCstep2}) is no more than linearly divergent 
so we need only the first term of (\ref{Dmexpansion}). The final 
contribution to (\ref{M3ptCstep2}) is,
\begin{equation}
\tfrac{D \kappa^2 m^4}{D-2} (a a')^D i\Delta_m(x;x') i\Delta_C(x;x') =
\tfrac{D \kappa^2 m^4}{D-2} \!\times\! \mathcal{K} a^4 i\delta^D(x \!-\! x')
+ {\rm Finite} \; . \label{Mdiv5}
\end{equation}
The key to reducing the other contributions on the last line is,
\begin{equation}
(a a')^{D-2} \partial_{\mu} i\Delta_m(x;x') \!\times\! i\Delta_C(x;x') =
\tfrac12 \mathcal{K} [\partial_{\mu} \!-\! (D\!-\! 2) a H \delta^0_{\mu}]
i\delta^D(x \!-\! x') + {\rm Finite} \; . \label{line2ID}
\end{equation}
After some partial integrations the result is,
\begin{eqnarray}
\lefteqn{\kappa^2 m^2 (a a')^{D-2} [{a'}^2 \overleftarrow{\partial} 
\!\!\cdot\!\! \overrightarrow{\partial} \!+\! a^2 \overleftarrow{\partial}' 
\!\!\cdot\!\! \overrightarrow{\partial}'] } \nonumber \\
& & \hspace{3cm} = -\tfrac12 \kappa^2 m^2 \mathcal{K} \Bigl\{
\overleftarrow{\mathcal{D}} \tfrac{i \delta^D(x - x')}{a^{D-4}} \!+\!
\tfrac{i \delta^D(x - x')}{a^{D-4}} \overrightarrow{\mathcal{D}}' \Bigr\}
+ {\rm Finite} \; . \qquad  \label{Mdiv4}
\end{eqnarray}

The terms on the first line of expression (\ref{M3ptCstep2}) are
quadratically divergent,
\begin{equation}
-\kappa^2 \overleftarrow{\partial}^{\mu} (a a')^{D-2} \Bigl\{ (\partial'_{\mu}
\partial_{\nu} \!+\! \eta_{\mu\nu} \partial \!\cdot\! \partial' \!-\! 
\partial_{\mu} \partial'_{\nu}) i\Delta_m(x;x') \!\times\! i\Delta_C(x;x') 
\Bigr\} {\overrightarrow{\partial}'}^{\nu} \; . \label{line1}
\end{equation}
These contributions requires the first two terms in the expansion
(\ref{Dmexpansion}) of the massive propagator,
\begin{eqnarray}
\lefteqn{ (a a')^{D-2} \partial_{\mu} \partial'_{\nu} i\Delta_m(x;x')
\!\times\! i\Delta_C(x;x') = \tfrac{\Gamma^2(\frac{D}2 -1)}{16 \pi^D} 
\Bigl\{ \tfrac{[\eta_{\mu\nu} \partial^2 - D \partial_{\mu} 
\partial_{\nu}]}{4 (D-1)} \tfrac1{\Delta x^{2D-4}} } \nonumber \\
& & \hspace{0cm} + \tfrac{(D-2) [a H \delta^0_{\mu} \partial_{\nu}
- a' H \delta^0_{\nu} \partial_{\mu}]}{4} \tfrac1{\Delta x^{2D-4}} 
+ \tfrac{(D-2)^2 a a' H^2 \delta^0_{\mu} \delta^0_{\nu}}{4 \Delta x^{2D-4}}
\nonumber \\
& & \hspace{6cm} - \tfrac{[m^2 - \frac{D}2 (\frac{D}2 - 1) H^2] a a' 
\eta_{\mu\nu}}{4\Delta x^{2D-4}} + {\rm Finite} \Bigr\} \; , \qquad \\
& & \hspace{-0.5cm} = \tfrac{\mathcal{K}}{4} \Bigl\{ \tfrac{[\eta_{\mu\nu} 
\partial^2 - D \partial_{\mu} \partial_{\nu}]}{D-1} \!+\! (D\!-\!2) [a H
\delta^0_{\mu} \partial_{\nu} \!-\! a' H \delta^0_{\nu} \partial_{\mu}]
\!+\! (D\!-\!2)^2 a a' H^2 \delta^0_{\mu} \delta^0_{\nu} \qquad 
\nonumber \\
& & \hspace{3cm} - [m^2 \!-\! \tfrac{D}2 (\tfrac{D-2}2 ) H^2] a a' 
\eta_{\mu\nu} \Bigr\} i\delta^D(x \!-\! x') + {\rm Finite} \; . \qquad
\label{line1ID}
\end{eqnarray}
Some judicious partial integrations imply,
\begin{eqnarray}
\lefteqn{ -\kappa^2 \overleftarrow{\partial}^{\mu} (a a')^{D-2} \Bigl\{ 
(\partial'_{\mu} \partial_{\nu} \!+\! \eta_{\mu\nu} \partial \!\cdot\! 
\partial' \!-\! \partial_{\mu} \partial'_{\nu}) i\Delta_m(x;x') \!\times\! 
i\Delta_C(x;x') \Bigr\} {\overrightarrow{\partial}'}^{\nu} } \nonumber \\
& & \hspace{-0.5cm} = \tfrac{\kappa^2 \mathcal{K}}{4} \Bigl\{ (D\!-\!2) 
\partial^{\mu} \partial'_0 (a \!+\! a') H \partial_{\mu} \!+\! (D \!-\!2)
{\partial'}^{\nu} \partial_0 (a \!+\! a') H \partial'_{\nu} \nonumber \\
& & \hspace{0cm} + \partial \!\cdot\! \partial' \Bigl[ {\scriptstyle 
(D- 2) (D-3) } H^2 \!-\! \tfrac{D^2}{2} (\tfrac{D-2}{2}) H^2 \!+\! D m^2 
\Bigr] a^2 \Bigr\} i\delta^D(x \!-\! x') + {\rm Finite} \; , \qquad \\
& & \hspace{-0.5cm} = \tfrac{\kappa^2 \mathcal{K}}{4} \Bigl\{ 
{\scriptstyle 2(D-2)} a^2 H^2 \nabla^2 i\delta^D(x \!-\! x') \nonumber \\
& & \hspace{1.7cm} - \Bigl[ {\scriptstyle (D- 2) (D-3) } H^2 \!-\! 
\tfrac{D^2}{2} (\tfrac{D-2}{2}) H^2 \!+\! D m^2 \Bigr] \mathcal{D} 
\tfrac{i \delta^D(x - x')}{a^{D-4}} \Bigr\} \!+\! {\rm Finite} . 
\qquad \label{Mdiv123}
\end{eqnarray} 

Although the contraction structure is the same as that of Diagram 7,
the presence of only two internal propagators means that the $i\Delta_{AC}$
terms contribute. The initial contraction produces three sorts of terms,
\begin{eqnarray}
\lefteqn{-i M^2_{\rm 3ptAC} = -\kappa^2 \Bigl\{ [
\overleftarrow{\nabla} \!\!\cdot\!\! \overleftarrow{\nabla}' \overrightarrow{\nabla}
\!\!\cdot\!\! \overrightarrow{\nabla}' \!+\! \overleftarrow{\nabla} \!\!\cdot\!\!
\overrightarrow{\nabla}' \overrightarrow{\nabla} \!\!\cdot\!\! 
\overleftarrow{\nabla}' \!-\! \tfrac{2}{D-3} \overleftarrow{\nabla} \!\!\cdot\!\!
\overrightarrow{\nabla} \overleftarrow{\nabla}' \!\!\cdot\!\! 
\overrightarrow{\nabla}' ] (a a')^{D-2} } \nonumber \\
& & \hspace{0cm} + \tfrac{2 a^{D-2}}{D-3} \overleftarrow{\nabla} \!\!\cdot\!\!
\overrightarrow{\nabla} (\overleftarrow{\partial}' {a'}^{D-2} \!\!\cdot\!\!
\overrightarrow{\partial}' \!+\! {a'}^D m^2) \!+\! \tfrac{2 {a'}^{D-2}}{D-3}
\overleftarrow{\nabla}' \!\!\cdot\!\! \overrightarrow{\nabla}'
(\overleftarrow{\partial} a^{D-2} \!\!\cdot\!\! \overrightarrow{\partial} \!+\! 
a^D m^2) \nonumber \\
& & \hspace{0.5cm} - (\tfrac{D-1}{D-3}) (\overleftarrow{\partial} a^{D-2} 
\!\!\cdot\!\! \overrightarrow{\partial} \!+\! a^D m^2) (\overleftarrow{\partial}' 
{a'}^{D-2} \!\!\cdot\!\! \overrightarrow{\partial}' \!+\! {a'}^D m^2) \Bigr\}
i\Delta_m \!\times\! i\Delta_{AC} \; . \qquad \label{M3ptAC}
\end{eqnarray}
Because at most two derivatives act on the internal propagators, we need only
retain the first terms in the expansion (\ref{Dmexpansion}) of $i\Delta_m$.
We also need only the most singular part of the expansion of $i\Delta_{AC}$,
\begin{equation}
i\Delta_{AC}(x;x') = \tfrac{H^2}{8 \pi^{\frac{D}2}} \tfrac{\Gamma(\frac{D}2 +1)}{
D-4} \tfrac1{(a a' \Delta x^2)^{\frac{D}2 - 2}} + \tfrac{k \Psi_A}{(a a')^{
\frac{D}2 - 2}} + {\rm Finite} \; . \label{DACexpansion}
\end{equation}
Note that the $n=0$ part of $i\Delta_C$ cannot contribute a divergence although
it does make a nonzero, finite contribution.

The key relation for the three terms on the first line of (\ref{M3ptAC}) is,
\begin{equation}
(a a')^{D-2} \partial_i \partial_j' i\Delta_m(x;x') \!\times\! i\Delta_{AC}(x;x') 
= \tfrac{D (D-2)}{16} a^2 H^2 \mathcal{K} i\delta^D(x \!-\! x') \delta_{ij} 
+ {\rm Finite} \; . \label{keyAC1}
\end{equation}
Hence the terms on the first line contribute,
\begin{eqnarray}
\lefteqn{-\kappa^2 \overleftarrow{\partial}_i (a a')^{D-2} \Bigl[
\partial_i' \partial_j \!+\! \delta_{ij} \vec{\nabla} \!\cdot\! \vec{\nabla}'
\!-\! \tfrac{2}{D-3} \partial_i \partial_j'\Bigr] i\Delta_m(x;x') \!\times\!
i\Delta_{AC}(x;x') \!\times\! \overrightarrow{\partial}_j' } \nonumber \\
& & \hspace{4.4cm} = \tfrac{D (D-2)^2}{16} \, \kappa^2 H^2 \mathcal{K} a^2
\nabla^2 i\delta^D(x \!-\! x') + {\rm Finite} . \qquad \label{M3ptACline1}
\end{eqnarray}
The first of the second line terms gives,
\begin{eqnarray}
\lefteqn{ -\tfrac{2 \kappa^2}{D-3} \overleftarrow{\partial}_i \Bigl\{
(a a')^{D-2} \partial_i {\partial'}^{\mu} i\Delta_m \!\times\!
i\Delta_{AC} \overrightarrow{\partial}'_{\mu} \!+\! m^2 a^{D-2} 
{a'}^D \partial_i i\Delta_m \!\times\! i\Delta_{AC} \Bigr\} } 
\nonumber \\
& & \hspace{4.5cm} = \tfrac{D (D-2)}{8 (D-3)} \kappa^2 H^2 \mathcal{K} 
a^2 \nabla^2 i\delta^D(x \!-\! x') + {\rm Finite} . \qquad
\label{M3ptACline2i}
\end{eqnarray}
The second term on line 2 makes the same divergent contribution,
\begin{eqnarray}
\lefteqn{ -\tfrac{2 \kappa^2}{D-3}  \Bigl\{\overleftarrow{\partial}^{\mu} 
(a a')^{D-2} \partial'_i \partial_{\mu} i\Delta_m \!\times\! i\Delta_{AC} 
\!+\! m^2 a^{D} {a'}^{D-2} \partial'_i i\Delta_m \!\times\! i\Delta_{AC} 
\Bigr\} \overrightarrow{\partial}'_i } \nonumber \\
& & \hspace{4.5cm} = \tfrac{D (D-2)}{8 (D-3)} \kappa^2 H^2 \mathcal{K} 
a^2 \nabla^2 i\delta^D(x \!-\! x') + {\rm Finite} . \qquad
\label{M3ptACline2ii}
\end{eqnarray}
And the final line of expression (\ref{M3ptAC}) gives,
\begin{eqnarray}
\lefteqn{ \kappa^2 (\tfrac{D-1}{D-3}) (\overleftarrow{\partial} a^{D-2} 
\!\!\cdot\!\! \overrightarrow{\partial} \!+\! a^D m^2) (\overleftarrow{\partial}' 
{a'}^{D-2} \!\!\cdot\!\! \overrightarrow{\partial}' \!+\! {a'}^D m^2)
i\Delta_m \, i\Delta_{AC} } \nonumber \\
& & \hspace{4.5cm} = -\tfrac{D (D-1) (D-2)}{16 (D-3)} \kappa^2 H^2
\mathcal{K} \mathcal{D} \tfrac{i \delta^D(x - x')}{a^{D-4}} \!+\! {\rm Finite} 
. \qquad \label{M3ptACline3}
\end{eqnarray}

The trace terms of the $i\Delta_{BC}$ contribution to $-iM^2_{\rm 3pt}(x;x')$ 
all vanish to give,
\begin{eqnarray}
\lefteqn{-i M^2_{\rm 3ptBC}(x;x') = \kappa^2 \Bigl\{ 
\overleftarrow{\partial}_0 (a a')^{D-2} \vec{\nabla} \!\cdot\! \vec{\nabla}'
i\Delta_m(x;x') \!\times\! i\Delta_{BC}(x;x') \overrightarrow{\partial}'_0 } 
\nonumber \\
& & \hspace{-0.5cm} + \overleftarrow{\partial}_0 (a a')^{D-2} \partial_0'
\partial_i i\Delta_m(x;x') \!\times\! i\Delta_{BC}(x;x') 
\overrightarrow{\partial}'_i + \overleftarrow{\partial}_i (a a')^{D-2} 
\partial_0 \partial'_i i\Delta_m(x;x') \nonumber \\
& & \hspace{0.5cm} \times i\Delta_{BC}(x;x') \overrightarrow{\partial}'_0 
\!+\! \overleftarrow{\partial}_i (a a')^{D-2} \partial_0 \partial_0'
i\Delta_m(x;x') \!\times\! i\Delta_{BC}(x;x') \overrightarrow{\partial}_i'
\Bigr\} . \qquad \label{M3ptBC}
\end{eqnarray}
It turns out that each of the four terms in expression (\ref{M3ptBC}) is
finite. To see this, use just the leading term in the expansion 
(\ref{Dmexpansion}) of the massive propagator and the leading term in the 
expansion of $i \Delta_{BC}$,
\begin{equation}
i\Delta_{BC}(x;x') = \tfrac{H^2}{8 \pi^{\frac{D}2}} 
\tfrac{\Gamma(\frac{D}2 - 1)}{(a a' \Delta x^2)^{\frac{D}2 - 2}} - 
\tfrac{k}{D-3} + {\rm Finite} \; . \label{DBCexpansion}
\end{equation}
It will suffice to consider the last of the four terms in expression
(\ref{M3ptBC}),
\begin{eqnarray}
\lefteqn{ (a a')^{D-2} \partial_0 \partial_0' i\Delta_m(x;x') \!\times\!
i\Delta_{BC}(x;x') = \tfrac{a a' H^2 \Gamma^2(\frac{D}2 -1)}{32 \pi^D} 
\Bigl\{ -\tfrac{(D-4)}{2 \Delta x^{2D -4}} } \nonumber \\
& & \hspace{1.2cm} + \tfrac14 (\tfrac{D}{D-3})
\partial_0 \partial_0' \tfrac1{\Delta x^{2D-6}} \Bigr\} - 
\tfrac{k (a a')^{\frac{D}2 -1}}{D-3} \tfrac{\Gamma(\frac{D}2 -1)}{4 
\pi^{\frac{D}2}} \partial_0 \partial_0' \tfrac1{\Delta x^{D-2}} 
+ {\rm Finite} \; . \qquad
\end{eqnarray}
This expression is finite. The same considerations apply to the other 
three terms in expression (\ref{M3ptBC}), so $-i M^2_{\rm 3ptBC}(x;x')$ 
contributes no divergence.

\subsection{Renormalization}

The divergent contributions to $-i M^2(x;x')$ reside on expressions
(\ref{M4div}), (\ref{Mdiv5}), (\ref{Mdiv4}), (\ref{Mdiv123}), and 
(\ref{M3ptACline1}-\ref{M3ptACline3}). Dropping some finite terms,
the sum is,
\begin{eqnarray}
\lefteqn{-i M_{\rm div}^2(x;x') = -\kappa^2 [m^2 \mathcal{K} \!+\! 
\tfrac32 k \Psi_A ] \Bigl\{ (\mathcal{D} \!-\! a^D m^2) \!+\! 
(\mathcal{D}' \!-\! {a'}^D m^2) \Bigr\} \tfrac{i \delta^D(x - x')}{a^{D-4}} 
} \nonumber \\
& & \hspace{6cm} + 4 \kappa^2 [H^2 \mathcal{K} \!+\! k \Psi_A] 
a^{D-2} \nabla^2 \tfrac{i \delta^D(x - x')}{a^{D-4}} . \qquad 
\label{divtot}
\end{eqnarray}
Recall that the leading forms of the various symbols are,
\begin{equation}
k = \tfrac{H^2}{8 \pi^2} + \dots \qquad , \qquad \mathcal{K} = \tfrac1{8\pi^2}
\tfrac1{D-4} + \dots \qquad , \qquad \Psi_A = -\tfrac{2}{D-4} + \dots \; . 
\label{constants}
\end{equation}

The 1-loop counterterms for the massive scalar self-mass are,
\begin{eqnarray}
\lefteqn{ \Delta \mathcal{L} = \tfrac12 c_3 (\square \!-\! m^2) \psi
(\square \!-\! m^2) \psi \sqrt{-g} - \tfrac12 c_4 [\partial_{\mu}
\psi \partial_{\nu} \psi g^{\mu\nu} \!+\! m^2 \psi^2] \sqrt{-g} }
\nonumber \\
& & \hspace{5.5cm} - \tfrac12 c_5 \psi^2 \sqrt{-g} \!-\! \tfrac12 c_6
R \partial_i \psi \partial_j \psi g^{ij} \sqrt{-g} \; . \qquad 
\label{DeltaLM}
\end{eqnarray}
Comparison with expression (\ref{divtot}) implies that the coefficients
are,
\begin{eqnarray}
c_3 = 0 \qquad & , & \qquad c_4 = \kappa^2 [2 m^2 \mathcal{K} \!+\! 3 k 
\Psi_A] \; , \qquad \label{c45} \\
c_5 = 0 \qquad & , & \qquad c_6 = -\tfrac13 \kappa^2 [\mathcal{K} \!+\!
\tfrac{k \Psi_A}{H^2}] \; . \qquad \label{c67}
\end{eqnarray}
The finite secular terms after renormalization are,
\begin{eqnarray}
\lefteqn{-i M^2_{\rm sec}(x;x') = \tfrac{\kappa^2 (m^2 - 3 H^2)}{
8 \pi^2} \Bigl\{ (\mathcal{D} \!-\! a^4 m^2) \!+\! (\mathcal{D}' \!-\!
{a'}^4 m^2) \Bigr\} \ln(a) i\delta^4(x \!-\! x') } \nonumber \\
& & \hspace{6.5cm} + \tfrac{\kappa^2 H^2}{2 \pi^2} \!\times\! a^2 \nabla^2 
[\ln(a) i\delta^4(x \!-\! x')] \; . \qquad \label{MrenI}
\end{eqnarray}
It is useful to re-express the $\nabla^2$ term using 4-point identities
(\ref{4ptID2}-\ref{4ptID3}),
\begin{eqnarray}
\lefteqn{-i M^2_{\rm sec}(x;x') = \tfrac{\kappa^2 (m^2 - 3 H^2)}{8 \pi^2} 
\Bigl\{ (\mathcal{D} \!-\! a^4 m^2) \!+\! (\mathcal{D}' \!-\!
{a'}^4 m^2) \Bigr\} \ln(a) i\delta^4(x \!-\! x') } \nonumber \\
& & \hspace{-0.5cm} + \tfrac{\kappa^2 H^2}{4 \pi^2} \Bigl\{ \! (\mathcal{D} 
\!-\! a^4 \! m^2) \ln(a) [\tfrac{\partial_0}{a H} \!+\! \tfrac32] \!+\! 
(\mathcal{D}' \!-\! {a'}^4 \!m^2) \ln(a') [\tfrac{\partial'_0}{a' H} \!+\! 
\tfrac32 ] \! \Bigr\} i\delta^4(x \!-\! x') . \qquad \label{MrenII}
\end{eqnarray}
The sense of this relation is that the operators $\mathcal{D}$ and 
$\mathcal{D}'$ on the second line act on all functions standing to their 
right.

\subsection{Correcting the Wavefunction}

The source-free effective field equation is,
\begin{equation}
(\mathcal{D} \!-\! a^4 m^2) \psi(x) = \int \!\! d^4x' \, M^2(x;x') \psi(x')
\; . \label{EFE}
\end{equation}
Because we only possess the 1-loop self-mass, this equation must be solved
perturbatively,
\begin{equation}
\psi = \psi_0 + \psi_1 + \dots \; . \label{Phiexp}
\end{equation}
The tree order and 1-loop equations are,
\begin{eqnarray}
(\mathcal{D} \!-\! a^4 m^2) \psi_0(x) &\!\!\! = \!\!\!& 0 \; , \label{tree} \\
(\mathcal{D} \!-\! a^4 m^2) \psi_1(x) &\!\!\! = \!\!\!& \int \!\! d^4x' \,
M^2_{1}(x;x') \psi_0(x') \; . \qquad \label{1loop}
\end{eqnarray}
Were the nonlocal, nonsecular parts of the self-mass included we would 
employ the Schwinger-Keldysh formalism \cite{Chou:1984es,Jordan:1986ug,
Calzetta:1986ey}, but this is not necessary to capture only the most 
important secular corrections. 

Because the factors of $(\mathcal{D}' - {a'}^4 m^2)$ in expression 
(\ref{MrenII}) can be partially integrated to annihilate the $\psi_0(x')$,
the secular parts of the 1-loop equation (\ref{1loop}) are,
\begin{eqnarray}
\lefteqn{(\mathcal{D} \!-\! a^4 m^2) \psi_1 = (\mathcal{D} \!-\! a^4 m^2)
} \nonumber \\
& & \hspace{0.7cm} \times \Bigl\{ -\tfrac{\kappa^2 (m^2 - 3 H^2)}{8 \pi^2} 
\ln(a) \psi_0 \!-\! \tfrac{\kappa^2 H^2}{4 \pi^2} \ln(a) 
[\tfrac{\partial_0}{a H} \!+\! \tfrac32] \psi_0\Bigr\} + {\rm Nonsecular} 
. \qquad \label{Phi1eqn}
\end{eqnarray}
The secularly enhanced part of the solution is easy to read off from
equation (\ref{Phi1eqn}),
\begin{eqnarray}
\psi_1 &\!\!\!\ = \!\!\!& -\tfrac{\kappa^2 (m^2 - 3 H^2)}{8 \pi^2} \ln(a) 
\psi_0 \!-\! \tfrac{\kappa^2 H^2}{4 \pi^2} \ln(a) [\tfrac{\partial_0}{a H} 
\!+\! \tfrac32] \psi_0 + {\rm Nonsecular} \; , \qquad \\
&\!\!\! = \!\!\!& -\tfrac{\kappa^2 m^2}{8 \pi^2} \ln(a) \psi_0 + 
{\rm Nonsecular} \; , \qquad \label{Phi1}
\end{eqnarray}
where we have neglected $(\frac{\partial_0}{a H} + \frac32) \psi_0$, which 
tends to cancel between in-coming and out-going legs. Each of the four legs 
is subject to this correction so the induced self-mass is,
\begin{equation}
-i M^2_{\delta u}(x;x') = \tfrac{\kappa^2 (m^2 - 3 H^2)}{4 \pi^2} 
\Bigl\{ \mathcal{D}_x [\ln(a_x) i\delta^4(x \!-\! x')] \!+\! \mathcal{D}_{x'} 
[\ln(a_{x'}) i\delta^4(x \!-\! x')] \Bigr\} . \label{Mdeltau}
\end{equation}

\section{Conclusions}

On flat space background it is by now well known how to combine bits and 
pieces of off-shell scattering amplitudes to infer gauge-independent graviton 
loop corrections to 1PI 2-point functions \cite{Miao:2017feh,Katuwal:2021thy}.
This procedure was recently extended to de Sitter background for the self-mass
of a massless, minimally coupled scalar \cite{Glavan:2024elz}. However, an
explicit check of gauge dependence revealed that the de Sitter computation
requires two new classes of diagrams and external state corrections that were 
not necessary in flat space \cite{Glavan:2026pug}. The point of this work has 
been to include those three additional corrections. The secular logarithm 
part of the original, uncorrected result \cite{Glavan:2021adm} (shown in 
Figure~\ref{Diagram0}) was given in expression (\ref{M0logs}), and is 
repeated here,
\begin{equation}
-i M^2_{0}(x;x') = -\tfrac{\kappa^2 H^2}{4 \pi^2} \partial_{x}^{\mu}
[ a_x^2 \ln(a_x) \partial^x_{\mu} i\delta^4(x \!-\! x')] + 
\tfrac{\kappa^2 H^2}{2 \pi^2} \partial_0^x [a_x^2 \ln(a_x) \partial^x_0
i\delta^4(x \!-\! x')] \; . \label{original}
\end{equation}
The de Sitter generalization of the five diagrams 
(Figures~\ref{Diagram1}-\ref{Diagram5}) which are needed to correct
the flat space result \cite{Glavan:2024elz} were reviewed in expressions 
(\ref{M4bar}-\ref{M1bc}) and contribute,
\begin{eqnarray}
\lefteqn{-i M^2_{1,2,3,4,5}(x;x') = -\tfrac{3 \kappa^2 H^2}{\pi^2} 
\Bigl\{\mathcal{D}_{x} [\ln(a_x) i\delta^4(x \!-\! x')] \!+\! \mathcal{D}_{x'} 
[\ln(a_{x'}) i\delta^4(x \!-\! x')] \Bigr\} } \nonumber \\
& & \hspace{7cm} + \tfrac{\kappa^2}{2 \pi^2} \mathcal{D}_{x} 
\mathcal{D}_{x'} \Bigl\{ \tfrac{\ln(a_x) i\delta^4(x - x')}{a_x^4} 
\Bigr\} \; . \qquad \label{1stcorrection}
\end{eqnarray}
Our results for the three extra corrections (Figures~\ref{Diagram6}-\ref{Diagram7}
and the wavefunction correction of Section 5) are given in expressions 
(\ref{Msq6total}), (\ref{M7}) and (\ref{Mdeltau}), whose sum is,
\begin{eqnarray}
\lefteqn{-i M^2_{\rm extra}(x;x') = -\tfrac{3 \kappa^2 H^2}{4\pi^2} 
\Bigl\{\mathcal{D}_{x} [\ln(a_x) i\delta^4(x \!-\! x')] \!+\! \mathcal{D}_{x'} 
[\ln(a_{x'}) i\delta^4(x \!-\! x')] \Bigr\} } \nonumber \\
& & \hspace{7cm} - \tfrac{\kappa^2}{8 \pi^2} \mathcal{D}_{x} 
\mathcal{D}_{x'} \Bigl\{ \tfrac{\ln(a_x) i\delta^4(x - x')}{a_x^4} 
\Bigr\} \; . \qquad \label{2ndcorrection}
\end{eqnarray}
The fact that expressions (\ref{original}-\ref{2ndcorrection}) have a
nonzero sum implies that large secular logarithmic corrections to the 
massless, minimally coupled scalar self-mass on de Sitter are not gauge 
artifacts. And we have already shown that these secular logarithms induce
a spatial logarithmic correction to the 1-loop exchange potential
\cite{Glavan:2021adm}. 

This has been a long and tedious exercise, and it is well to comment on
the checks we have on its accuracy. The first check is that each of the
three extra corrections agrees with its flat space limit when the de Sitter
Hubble parameter goes to zero. Another important check is that each of the
three extra contributions --- (\ref{Msq6total}), (\ref{M7}) and 
(\ref{Mdeltau}) --- contains a term proportional to $m^2$, yet these 
cancel in the sum (\ref{2ndcorrection}),
\begin{equation}
-\tfrac{3 \kappa^2 m^2}{8\pi^2} + \tfrac{\kappa^2 m^2}{8 \pi^2}
+ \tfrac{\kappa^2 m^2}{4 \pi^2} = 0 \; . \label{Mcancellation}
\end{equation}
Had the $m^2$ terms failed to cancel, it would have meant that extra 
diagrams need to be included. The final check concerns divergences in
the massive scalar self-mass (\ref{divtot}). Note that the $-\frac12 c_5
\psi^2 \sqrt{-g}$ counterterm in expression (\ref{DeltaLM}) vanishes.
That is why equation (\ref{Phi1eqn}) was so easy to solve. Had the
coefficient $c_5$ been nonzero, the 1-loop solution would go like $\ln^2(a)
\times \Phi_0$, which would completely dominate the other contributions,
(\ref{original}-\ref{2ndcorrection}).

Besides establishing that graviton-induced logarithmic corrections are
real, a major implication of this work is that external legs can contribute
secular logarithms, even if the observer field $\psi$ is super-heavy. This
is quite different from flat space, where long-range effects require the
collaboration of {\it two} massless propagators \cite{Donoghue:1994dn,
Bjerrum-Bohr:2002aqa,Miao:2017feh,Katuwal:2021thy}; on de Sitter background
a single massless propagator will suffice. The reason for this seems to be
the ultraviolet origin of the graviton-induced logarithms (explained in 
Section 2.2) that have previously been found for massless fermions 
\cite{Miao:2005am}, for massless, minimally coupled scalars 
\cite{Kahya:2007bc}, and for photons \cite{Leonard:2013xsa}. These previous 
examples were all massless, but that turns out not to be necessary. Graviton 
loops can induce secular logarithms even in massive fields. That is why we 
refer to the external leg corrections studied in this paper as ``universal''.
 
Finally, we should comment on further work. The external leg corrections 
we have included here should suffice to remove gauge dependence from the 
self-mass of the massless, minimally coupled scalar. That is certainly true
of variations with respect to the parameter $\alpha$ in the de Sitter 
analog of (\ref{gauge}). It was checking this that revealed the need for
the three extra corrections we have included here \cite{Glavan:2026pug}. It 
remains to check for independence under variations of the parameter $\beta$. 
This should not reveal the need for any extra diagrams but it might serve 
to check the de Sitter generalizations of the ``Donoghue Identities'', 
analogous to relation (\ref{IntID1}), that were used to extract the 2-point 
contributions from 3-point and 4-point amplitudes. 

The other follow-up project is to remove gauge dependence from graviton
loop corrections to the fermion self-energy \cite{Miao:2005am}, the vacuum 
polarization \cite{Leonard:2013xsa,Glavan:2015ura}, and the graviton self-energy
\cite{Tsamis:1996qk}. The simplest of these is the vacuum polarization,
which has already been done in flat space \cite{Katuwal:2021thy}. We need
only take the de Sitter generalizations of the same flat space diagrams
that were used there, plus a handful of universal external leg corrections.
The massless fermion self-energy could probably be probed by coupling it 
to a massive fermion and a massive scalar, and then considering the 
process of massive fermion + massive scalar goes to massive scalar plus
massive fermion. The most difficult case is undoubtedly the graviton
self-energy. Not only does it involve two intermediate graviton lines,
it almost certainly manifests secular logarithms from the tail part of
the graviton propagator, in addition to the mismatch between primitive
divergences and counterterms that was described in Section 2.2. One 
indication of this is the fact that 1-loop corrections to the Newtonian 
potential on de Sitter contain {\it three} secular logarithms, as opposed 
to the single logarithm that could come from ultraviolet effects 
\cite{Tan:2022xpn}.

\vskip 0.5cm 

\centerline{\bf Acknowledgements}

DG was supported by project 24-13079S of the Czech 
Science Foundation (GA\v{C}R). SPM was supported by Taiwan 
NSTC grants 113-2112-M-006-013 and 114-2112-M-006-020. TP is 
supported by the NWA ORC 2023 consortium grant: Cosmic emergence:
from abstract simplicity to complex diversity (Kosmische 
emergentie: van abstracte eenvoud naar complexe diversiteit).
TP and DG are funded by The Magnetic Universe NWO grant 
OCENW.XL.23.147. RPW was partially supported by NSF grant 
PHY-2207514 and by the Institute for Fundamental Theory at the 
University of Florida.


\begin{thebibliography}{99}

%\cite{Schwinger:1960qe}
\bibitem{Schwinger:1960qe}
J.~S.~Schwinger,
%``Brownian motion of a quantum oscillator,''
J. Math. Phys. \textbf{2}, 407-432 (1961)
doi:10.1063/1.1703727
%2148 citations counted in INSPIRE as of 20 Feb 2026

%\cite{Mahanthappa:1962ex}
\bibitem{Mahanthappa:1962ex}
K.~T.~Mahanthappa,
%``Multiple production of photons in quantum electrodynamics,''
Phys. Rev. \textbf{126}, 329-340 (1962)
doi:10.1103/PhysRev.126.329
%345 citations counted in INSPIRE as of 20 Feb 2026

%\cite{Bakshi:1962dv}
\bibitem{Bakshi:1962dv}
P.~M.~Bakshi and K.~T.~Mahanthappa,
%``Expectation value formalism in quantum field theory. 1.,''
J. Math. Phys. \textbf{4}, 1-11 (1963)
doi:10.1063/1.1703883
%419 citations counted in INSPIRE as of 20 Feb 2026

%\cite{Bakshi:1963bn}
\bibitem{Bakshi:1963bn}
P.~M.~Bakshi and K.~T.~Mahanthappa,
%``Expectation value formalism in quantum field theory. 2.,''
J. Math. Phys. \textbf{4}, 12-16 (1963)
doi:10.1063/1.1703879
%374 citations counted in INSPIRE as of 20 Feb 2026

%\cite{Keldysh:1964ud}
\bibitem{Keldysh:1964ud}
L.~V.~Keldysh,
%``Diagram Technique for Nonequilibrium Processes,''
Sov. Phys. JETP \textbf{20}, 1018-1026 (1965)
doi:10.1142/9789811279461{\_}0007
%2264 citations counted in INSPIRE as of 20 Feb 2026

%\cite{Chou:1984es}
\bibitem{Chou:1984es}
K.~c.~Chou, Z.~b.~Su, B.~l.~Hao and L.~Yu,
%``Equilibrium and Nonequilibrium Formalisms Made Unified,''
Phys. Rept. \textbf{118}, 1-131 (1985)
doi:10.1016/0370-1573(85)90136-X
%995 citations counted in INSPIRE as of 20 Feb 2026

%\cite{Jordan:1986ug}
\bibitem{Jordan:1986ug}
R.~D.~Jordan,
%``Effective Field Equations for Expectation Values,''
Phys. Rev. D \textbf{33}, 444-454 (1986)
doi:10.1103/PhysRevD.33.444
%493 citations counted in INSPIRE as of 20 Feb 2026

%\cite{Calzetta:1986ey}
\bibitem{Calzetta:1986ey}
E.~Calzetta and B.~L.~Hu,
%``Closed Time Path Functional Formalism in Curved Space-Time: Application to Cosmological Back Reaction Problems,''
Phys. Rev. D \textbf{35}, 495 (1987)
doi:10.1103/PhysRevD.35.495
%557 citations counted in INSPIRE as of 20 Feb 2026

%\cite{Ford:2004wc}
\bibitem{Ford:2004wc}
L.~H.~Ford and R.~P.~Woodard,
%``Stress tensor correlators in the Schwinger-Keldysh formalism,''
Class. Quant. Grav. \textbf{22}, 1637-1647 (2005)
doi:10.1088/0264-9381/22/9/011
[arXiv:gr-qc/0411003 [gr-qc]].
%74 citations counted in INSPIRE as of 20 Feb 2026

%\cite{Miao:2017feh}
\bibitem{Miao:2017feh}
S.~P.~Miao, T.~Prokopec and R.~P.~Woodard,
%``Deducing Cosmological Observables from the S-matrix,''
Phys. Rev. D \textbf{96}, no.10, 104029 (2017)
doi:10.1103/PhysRevD.96.104029
[arXiv:1708.06239 [gr-qc]].
%28 citations counted in INSPIRE as of 03 Feb 2026

%\cite{Donoghue:1993eb}
\bibitem{Donoghue:1993eb}
J.~F.~Donoghue,
%``Leading quantum correction to the Newtonian potential,''
Phys. Rev. Lett. \textbf{72}, 2996-2999 (1994)
doi:10.1103/PhysRevLett.72.2996
[arXiv:gr-qc/9310024 [gr-qc]].
%654 citations counted in INSPIRE as of 13 Jul 2026

%\cite{Donoghue:1994dn}
\bibitem{Donoghue:1994dn}
J.~F.~Donoghue,
%``General relativity as an effective field theory: The leading quantum corrections,''
Phys. Rev. D \textbf{50}, 3874-3888 (1994)
doi:10.1103/PhysRevD.50.3874
[arXiv:gr-qc/9405057 [gr-qc]].
%1569 citations counted in INSPIRE as of 13 Jul 2026

%\cite{Donoghue:1996mt}
\bibitem{Donoghue:1996mt}
J.~F.~Donoghue and T.~Torma,
%``On the power counting of loop diagrams in general relativity,''
Phys. Rev. D \textbf{54}, 4963-4972 (1996)
doi:10.1103/PhysRevD.54.4963
[arXiv:hep-th/9602121 [hep-th]].
%75 citations counted in INSPIRE as of 13 Jul 2026

%\cite{Katuwal:2021thy}
\bibitem{Katuwal:2021thy}
S.~Katuwal and R.~P.~Woodard,
%``Gauge independent quantum gravitational corrections to Maxwell{\textquoteright}s equation,''
JHEP \textbf{10}, 029 (2021)
doi:10.1007/JHEP10(2021)029
[arXiv:2107.13341 [gr-qc]].
%8 citations counted in INSPIRE as of 03 Feb 2026

%\cite{Glavan:2019msf}
\bibitem{Glavan:2019msf}
D.~Glavan, S.~P.~Miao, T.~Prokopec and R.~P.~Woodard,
%``Graviton Propagator in a 2-Parameter Family of de Sitter Breaking Gauges,''
JHEP \textbf{10}, 096 (2019)
doi:10.1007/JHEP10(2019)096
[arXiv:1908.06064 [gr-qc]].
%16 citations counted in INSPIRE as of 08 Sep 2026

%\cite{Glavan:2021adm}
\bibitem{Glavan:2021adm}
D.~Glavan, S.~P.~Miao, T.~Prokopec and R.~P.~Woodard,
%``Large logarithms from quantum gravitational corrections to a massless, minimally coupled scalar on de Sitter,''
JHEP \textbf{03}, 088 (2022)
doi:10.1007/JHEP03(2022)088
[arXiv:2112.00959 [gr-qc]].
%30 citations counted in INSPIRE as of 03 Feb 2026

%\cite{Glavan:2024elz}
\bibitem{Glavan:2024elz}
D.~Glavan, S.~P.~Miao, T.~Prokopec and R.~P.~Woodard,
%``Gauge independent logarithms from inflationary gravitons,''
JHEP \textbf{03}, 129 (2024)
doi:10.1007/JHEP03(2024)129
[arXiv:2402.05452 [hep-th]].
%6 citations counted in INSPIRE as of 03 Feb 2026

%\cite{Glavan:2025azq}
\bibitem{Glavan:2025azq}
D.~Glavan,
%``Graviton propagator in de Sitter space in a simple one-parameter gauge,''
JHEP \textbf{05}, 189 (2026)
doi:10.1007/JHEP05(2026)189
[arXiv:2511.13660 [gr-qc]].
%4 citations counted in INSPIRE as of 08 Sep 2026

%\cite{Glavan:2026pug}
\bibitem{Glavan:2026pug}
D.~Glavan, S.~P.~Miao, T.~Prokopec and R.~P.~Woodard,
%``Cancellation of one-parameter graviton gauge dependence in the effective scalar field equation in de Sitter,''
JHEP \textbf{04}, 159 (2026)
doi:10.1007/JHEP04(2026)159
[arXiv:2602.07908 [hep-th]].
%3 citations counted in INSPIRE as of 08 Sep 2026

%\cite{Burko:2002ge}
\bibitem{Burko:2002ge}
L.~M.~Burko, A.~I.~Harte and E.~Poisson,
%``Mass loss by a scalar charge in an expanding universe,''
Phys. Rev. D \textbf{65}, 124006 (2002)
doi:10.1103/PhysRevD.65.124006
[arXiv:gr-qc/0201020 [gr-qc]].
%50 citations counted in INSPIRE as of 08 Sep 2026

%\cite{Akhmedov:2010ah}
\bibitem{Akhmedov:2010ah}
E.~T.~Akhmedov, A.~Roura and A.~Sadofyev,
%``Classical radiation by free-falling charges in de Sitter spacetime,''
Phys. Rev. D \textbf{82}, 044035 (2010)
doi:10.1103/PhysRevD.82.044035
[arXiv:1006.3274 [gr-qc]].
%39 citations counted in INSPIRE as of 08 Sep 2026

%\cite{Glavan:2019yfc}
\bibitem{Glavan:2019yfc}
D.~Glavan, S.~P.~Miao, T.~Prokopec and R.~P.~Woodard,
%``Breaking of scaling symmetry by massless scalar on de Sitter,''
Phys. Lett. B \textbf{798}, 134944 (2019)
doi:10.1016/j.physletb.2019.134944
[arXiv:1908.11113 [gr-qc]].
%10 citations counted in INSPIRE as of 08 Sep 2026

%\cite{Chernikov:1968zm}
\bibitem{Chernikov:1968zm}
N.~A.~Chernikov and E.~A.~Tagirov,
%``Quantum theory of scalar field in de Sitter space-time,''
Ann. Inst. H. Poincare Phys. Theor. A \textbf{9}, no.2, 109-141 (1968)
%396 citations counted in INSPIRE as of 01 Mar 2026

%\cite{Schomblond:1976xc}
\bibitem{Schomblond:1976xc}
C.~Schomblond and P.~Spindel,
%``Unicity Conditions of the Scalar Field Propagator Delta(1) (x,y) in de Sitter Universe,''
Ann. Inst. H. Poincare Phys. Theor. \textbf{25}, 67-78 (1976)
%76 citations counted in INSPIRE as of 01 Mar 2026

%\cite{Bunch:1978yq}
\bibitem{Bunch:1978yq}
T.~S.~Bunch and P.~C.~W.~Davies,
%``Quantum Field Theory in de Sitter Space: Renormalization by Point Splitting,''
Proc. Roy. Soc. Lond. A \textbf{360}, 117-134 (1978)
doi:10.1098/rspa.1978.0060
%1430 citations counted in INSPIRE as of 01 Mar 2026

%\cite{Allen:1987tz}
\bibitem{Allen:1987tz}
B.~Allen and A.~Folacci,
%``The Massless Minimally Coupled Scalar Field in De Sitter Space,''
Phys. Rev. D \textbf{35}, 3771 (1987)
doi:10.1103/PhysRevD.35.3771
%382 citations counted in INSPIRE as of 02 Mar 2026

%\cite{Onemli:2002hr}
\bibitem{Onemli:2002hr}
V.~K.~Onemli and R.~P.~Woodard,
%``Superacceleration from massless, minimally coupled phi**4,''
Class. Quant. Grav. \textbf{19}, 4607 (2002)
doi:10.1088/0264-9381/19/17/311
[arXiv:gr-qc/0204065 [gr-qc]].
%429 citations counted in INSPIRE as of 02 Mar 2026

%\cite{Tsamis:1992xa}
\bibitem{Tsamis:1992xa}
N.~C.~Tsamis and R.~P.~Woodard,
%``The Structure of perturbative quantum gravity on a De Sitter background,''
Commun. Math. Phys. \textbf{162}, 217-248 (1994)
doi:10.1007/BF02102015
%168 citations counted in INSPIRE as of 02 Mar 2026

%\cite{Woodard:2004ut}
\bibitem{Woodard:2004ut}
R.~P.~Woodard,
%``de Sitter breaking in field theory,''
[arXiv:gr-qc/0408002 [gr-qc]].
%77 citations counted in INSPIRE as of 02 Mar 2026

%\cite{Woodard:2025smz}
\bibitem{Woodard:2025smz}
R.~P.~Woodard,
%``Resummations for inflationary quantum gravity,''
Int. J. Mod. Phys. D \textbf{34}, no.10, 2542002 (2025)
doi:10.1142/S0218271825420027
[arXiv:2501.05077 [gr-qc]].
%12 citations counted in INSPIRE as of 15 Jul 2026

%\cite{Bjerrum-Bohr:2002aqa}
\bibitem{Bjerrum-Bohr:2002aqa}
N.~E.~J.~Bjerrum-Bohr,
%``Leading quantum gravitational corrections to scalar QED,''
Phys. Rev. D \textbf{66}, 084023 (2002)
doi:10.1103/PhysRevD.66.084023
[arXiv:hep-th/0206236 [hep-th]].
%78 citations counted in INSPIRE as of 07 Aug 2026

%\cite{Miao:2005am}
\bibitem{Miao:2005am}
S.~P.~Miao and R.~P.~Woodard,
%``The Fermion self-energy during inflation,''
Class. Quant. Grav. \textbf{23}, 1721-1762 (2006)
doi:10.1088/0264-9381/23/5/016
[arXiv:gr-qc/0511140 [gr-qc]].
%128 citations counted in INSPIRE as of 07 Aug 2026

%\cite{Kahya:2007bc}
\bibitem{Kahya:2007bc}
E.~O.~Kahya and R.~P.~Woodard,
%``Quantum Gravity Corrections to the One Loop Scalar Self-Mass during Inflation,''
Phys. Rev. D \textbf{76}, 124005 (2007)
doi:10.1103/PhysRevD.76.124005
[arXiv:0709.0536 [gr-qc]].
%85 citations counted in INSPIRE as of 07 Aug 2026

%\cite{Leonard:2013xsa}
\bibitem{Leonard:2013xsa}
K.~E.~Leonard and R.~P.~Woodard,
%``Graviton Corrections to Vacuum Polarization during Inflation,''
Class. Quant. Grav. \textbf{31}, 015010 (2014)
doi:10.1088/0264-9381/31/1/015010
[arXiv:1304.7265 [gr-qc]].
%55 citations counted in INSPIRE as of 07 Aug 2026

%\cite{Glavan:2015ura}
\bibitem{Glavan:2015ura}
D.~Glavan, S.~P.~Miao, T.~Prokopec and R.~P.~Woodard,
%``Graviton Loop Corrections to Vacuum Polarization in de Sitter in a General Covariant Gauge,''
Class. Quant. Grav. \textbf{32}, no.19, 195014 (2015)
doi:10.1088/0264-9381/32/19/195014
[arXiv:1504.00894 [gr-qc]].
%37 citations counted in INSPIRE as of 08 Sep 2026

%\cite{Tsamis:1996qk}
\bibitem{Tsamis:1996qk}
N.~C.~Tsamis and R.~P.~Woodard,
%``One loop graviton selfenergy in a locally de Sitter background,''
Phys. Rev. D \textbf{54}, 2621-2639 (1996)
doi:10.1103/PhysRevD.54.2621
[arXiv:hep-ph/9602317 [hep-ph]].
%123 citations counted in INSPIRE as of 07 Aug 2026

%\cite{Tan:2022xpn}
\bibitem{Tan:2022xpn}
L.~Tan, N.~C.~Tsamis and R.~P.~Woodard,
%``How Inflationary Gravitons Affect the Force of Gravity,''
Universe \textbf{8}, no.7, 376 (2022)
doi:10.3390/universe8070376
[arXiv:2206.11467 [gr-qc]].
%19 citations counted in INSPIRE as of 07 Aug 2026

\end{thebibliography}
\end{document}